\documentclass[a4paper,fleqn,usenatbib]{mnras}

\usepackage{newtxtext,newtxmath}

\usepackage[T1]{fontenc}
\usepackage{ae,aecompl}

\usepackage{graphicx}	
\usepackage{amsmath}	
\usepackage{amssymb}	
\newcommand{\nd}{\multicolumn{1}{c}{$\dots$}}

\title[SN 2014J in M82]{SN 2014J in M82: New Insights On the Spectral Diversity of Type Ia Supernovae}

\author[K. Zhang et al.]{
Kaicheng Zhang,$^{1,2}$
Xiaofeng Wang,$^{1}$\thanks{E-mail: wang\_xf@mail.tsinghua.edu.cn}
JuJia Zhang,$^{3,4,5}$
Tianmeng Zhang,$^{6,7}$
S. Benetti,$^{8}$
\newauthor
N. Elias-Rosa,$^{8}$
Fang Huang,$^{9}$
Han Lin,$^{1}$
Linyi Li,$^{1}$
Wenxiong Li,$^{1}$
P. Ochner,$^{8,10}$
\newauthor
A. Pastorello,$^{8}$
Liming Rui,$^{1}$
L. Tartaglia,$^{11}$
L. Tomasella,$^{8}$
A. Siviero,$^{10}$
U. Munari,$^{8}$
\newauthor
G. Terreran,$^{12}$
Hao Song,$^{1}$
S. Taubenberger,$^{13,14}$
J. Craig Wheeler,$^{2}$
Danfeng Xiang,$^{1}$
\newauthor
Xulin Zhao,$^{15}$
Hongbin Li,$^{6}$
Jinming Bai,$^{3,4,5}$
Xiaojun Jiang,$^{6}$
Jianrong Shi,$^{6}$
Zhenyu Wu$^{6,7}$
\\
$^{1}$Physics Department and Tsinghua Center for Astrophysics (THCA), Tsinghua University, Beijing, 100084, China\\
$^{2}$Department of Astronomy, University of Texas at Austin, Austin, TX 78712-1205, USA\\
$^{3}$Yunnan Observatories, Chinese Academy of Sciences, Kunming, 650216, China\\
$^{4}$Key Laboratory for the Structure and Evolution of Celestial Objects, Chinese Academy of Sciences, Kunming, 650216, China\\
$^{5}$Center for Astronomical Mega-Science, Chinese Academy of Sciences, 20A Datun Road, Chaoyang District, Beijing, 100012, China\\
$^{6}$Key Laboratory of Optical Astronomy,National Astronomical Observatory of China, Chinese Academy of Sciences, Beijing, 100012, China\\
$^{7}$School of Astronomy and Space Science, University of Chinese Academy of Sciences, Beijing 101408, China.\\
$^{8}$INAF - Osservatorio Astronomico di Padova, vicolo dell'Osservatorio 5, I-35122 Padova, Italy\\
$^{9}$Department of Astronomy, School of Physics and Astronomy, Shanghai Jiao Tong University, Shanghai, 200240, China\\
$^{10}$Dipartimento di Fisica e Astronomia, Universit\`{a} di Padova, via Marzolo 8, I-35131 Padova, Italy\\
$^{11}$Department of Astronomy and The Oskar Klein Centre, AlbaNova
University Center, Stockholm University, SE-106 91 Stockholm, Sweden\\
$^{12}$Center for Interdisciplinary Exploration and Research in Astrophysics (CIERA) and Department of Physics and Astronomy, Northwestern University, Evanston, IL 60208, USA\\
$^{13}$European Southern Observatory, Karl-Schwarzschild-Str. 2, 85748 Garching, Germany\\
$^{14}$Max-Planck-Institut f\"ur Astrophysik, Karl-Schwarzschild-Str. 1, 85748 Garching, Germany\\
$^{15}$School of Science, Tianjin University of Technology, Tianjin, 300384, China
}

\date{Accepted XXX. Received YYY; in original form ZZZ}

\pubyear{2018}

\begin{document}
\label{firstpage}
\pagerange{\pageref{firstpage}--\pageref{lastpage}}
\maketitle

\begin{abstract}

We present extensive spectroscopic observations for one of the closest type Ia supernovae (SNe Ia), SN 2014J discovered in M82, ranging from 10.4 days before to 473.2 days after $B$-band maximum light. The diffuse interstellar band (DIB) features detected in a high-resolution spectrum allow an estimate of line-of-sight extinction as $A_\textrm{v}$ $\sim$1.9$\pm$0.6 mag. Spectroscopically, SN 2014J can be put into the high-velocity (HV) subgroup in Wang's classification with a velocity of Si~{\sc ii} $\lambda$ 6355 at maximum light of $v_0=1.22\pm 0.01 \times 10^4$ km~s$^{-1}$, but has a low velocity gradient (LVG, following Benetti's classification) of $\dot{v}=41\pm2$ km s$^{-1}$ day$^{-1}$, which is inconsistent with the trend that HV SNe Ia generally have larger velocity gradients. We find that the HV SNe Ia with LVGs tend to have relatively stronger Si~{\sc iii} (at $\sim$4400 \AA) absorptions in early spectra, larger ratios of S~{\sc ii}~$\lambda$ 5468 to S~{\sc ii}~$\lambda$ 5640, and weaker Si~{\sc ii} 5972 absorptions compared to their counterparts with similar velocities but high velocity gradients. This shows that the HV+LVG subgroup of SNe Ia may have intrinsically higher photospheric temperature, which indicates that their progenitors may experience more complete burning in the explosions relative to the typical HV SNe Ia.
\end{abstract}


\section{INTRODUCTION}
SN 2014J was discovered in the edge-on starburst galaxy M82 on 2014 January 21.805 (UT dates are used throughout this paper) by Fossey et al. (2014) and it was classified as a type Ia supernova (SN Ia) by Cao et al. (2014). SN 2014J is one of the nearest SNe Ia discovered over the past three decades, with a distance of only $\sim$3.5 Mpc. Extensive followup observations were obtained for this nearby SN Ia soon after its discovery.

SN 2014J reached its $B$-band maximum of 11.68$\pm$0.01 mag on MJD 56689.74$\pm$0.13 (Marion et al. 2015), with a post-peak decline rate as $\Delta$ m$_{15}$(B) as 0.96$\pm$0.03 mag (observed value from Srivastav et al. 2016). After correcting for the reddening effect (Phillips et al. 1999) from Milky Way and host galaxy, the true light-curve decline rate $\Delta m_{15}(B)_{true}$ is estimated as 1.08$\pm$0.03 (Srivastav et al. 2016). Based on the early photometric data, the explosion time was estimated to be 2014 Jan. 15.57 UT (MJD = 56672.57, Zheng et al. 2014).

With the near-infrared (NIR) and optical spectra from $-9.9$ d to +10.0 d, Marion et al. (2015) identified C~{\sc i} $\lambda$10693. They also found that SN 2014J has a layered structure with little or no mixing, which is consistent with delayed detonation explosion models (H{\"o}flich et al. 2002). Marion et al. (2015) and Srivastav et al. (2016), presenting spectra covering the phases from $-$7.71 d to +351.09 d, suggested that SN 2014J is near the border of the Normal Velocity (NV) group and the High Velocity (HV) group in the classification scheme of Wang et al. (2009a), belongs to the low-velocity gradient (LVG) subgroup in the classification scheme of Benetti et al. (2005), and lies at the border of the Core Normal (CN) and Broad Line (BL) subclasses in the classification scheme of Branch et al. (2009). Galbany et al. (2016) also concluded that SN 2014J is a transitional SN Ia in all the classification schemes.

Bright supernovae such as SN 2014J also provide good opportunities to study circumstellar material (CSM) and interstellar material (ISM) along the line of sight, given the relatively high extinction that it suffers. Using high-resolution spectra of SN 2014J in the early phase, Goobar et al. (2014) analyzed the dense intervening material and did not detect any evolution in the resolved absorption features during the rising phase of the light curves. In a series of highest resolution (R$\sim$110,000) spectra of SN 2014J, Graham et al. (2015; hereafter G15) did not detect evolution in any component of Na {\sc i} D and Ca {\sc ii}. However, they established the dissipation/weakening of the two most blueshifted components of K~{\sc i} lines, which was attributed to the photoionization of CSM, favoring a single-degenerate (SD) scenario for SN 2014J. The corresponding velocity components of Na {\sc i} D did not vary with time (which was also noticed by Goobar et al. 2014), but this may be due to its higher ionization energy. Ritchey et al. (2015) and Welty et al. (2014) detected the Na {\sc i}, Ca {\sc ii}, K {\sc i}, Ca {\sc i}, CH$^+$, CH, CN in the high-resolution spectra obtained for SN 2014J between $-$5.6 d and +30.4 d. In particular, the Li {\sc i} detected in the spectrum of SN 2014J is the first report for the detection of interstellar Li beyond the Local Group.

The high-resolution spectra of SN 2014J also allow the study of the diffuse interstellar bands (DIBs). It has been a long time since Heger (1922) first reported the DIBs, but the carriers of DIBs are still under debate. The earliest DIB feature detected in the spectra of a supernova was found by Rich (1987) using the spectra of SN 1986G. For SN 2014J, the DIB features have been studied by Welty et al. (2014), Goobar et al. (2014), and G15. Owing to the correlation with dust extinction, the DIB features can help determine the extinction of SN 2014J in its host galaxy M82.

In this paper, we present our extensive optical spectroscopic observations for SN 2014J as well as our analysis of the spectral features. We describe the observations and data reduction in \S 2. Analysis of the spectral features is presented in \S 3. We discussed the diversity of spectral features in \S 4, and we summarize our results in \S 5.

\section{DATA REDUCTION}

Our optical spectra of SN 2014J were obtained with the 2.16-m telescope at Xinglong Observatory of NAOC, the 2.4-m Lijiang Telescope of Yunnan Astronomical Observatory (YNAO + YFOSC), the 3.58-m Telescopio Nazionale Galileo (TNG + LRS), the Gran Telescopio Canarias (GTC+OSIRIS), the Copernico 1.82m telescope of the INAF-Padova Observatory, and the Galileo 1.22m telescope of the Padova University. Table 1 lists the spectroscopic observations.

We collected in total 56 optical spectra of SN 2014J, spanning from $t=-$10.4 days to $t=+$473.2 days with respect to the $B$-band maximum light (MJD 56689.74$\pm$0.13; Marion et al. 2015), which may represent the largest spectral dataset of SN 2014J published so far. Among our spectra, there are 6 intermediate-resolution spectra (with $R\approx$ 7,000-10,000) and 1 high-resolution spectrum (with $R\approx 60,000$) obtained with the Lijiang 2.4-m telescope and the Xinglong 2.16-m telescope, respectively. Strong narrow Na {\sc i} D absorptions are present in all of the spectra, indicating that SN 2014J suffered significant reddening from its host galaxy. Figure 1 shows the complete spectral evolution.

All of the low-resolution spectra obtained by these telescopes were reduced using standard IRAF routines, and calibrated with the help of spectrophotometric flux standard stars. The telluric lines in all the spectra are removed. In some of the 1.22-m telescope spectra, the removals of telluric lines were not perfect due to variable humidity and variable fringing patterns. Moreover, the strongest telluric absorption exactly matched the O {\sc i} 7774 absorption for nearby SNe, which makes the removals of the telluric lines more difficult. The Yunnan Faint Object Spectragraph Camera (YFOSC) mounted on the Lijiang 2.4-m telescope of YNAO is a multi-mode observation instrument, which can be used to take both low-resolution and intermediate-resolution spectra through cross dispersion. For example, the combination of G10 and E13 gratings can achieve a resolution of $R\approx$ 7,000-9,000, while the combination of G10 and E9 gratings can achieve a resolution of $R\approx$ 10,000. The high-resolution spectroscopy at the 2.16~m telescope at Xinglong Observatory is also achieved through cross dispersion. The cross dispersion spectra were reduced using the IRAF echelle package, including bias, flat, cosmic-ray correction, wavelength calibration and flux calibration. All the spectra were corrected for the continuum atmospheric extinction at the observatories where the spectra were obtained.

\section{SPECTRAL ANALYSIS}

\subsection{Spectra Comparison}

In Figure 2, we compare the spectra of SN 2014J from $\sim -10$ d to $\sim +9$ d with those of some normal SNe Ia with similar reddening-corrected decline rate (dubbed as $\Delta m_{15}(B)_{true}$), such as SNe 2002bo (Benetti et al. 2004), 2003du (Stanishev et al. 2007), 2005cf (Wang et al. 2009b), 2007co (Silverman et al. 2015), 2007le (Silverman et al. 2015), 2009ig (Marion et al. 2013), and 2011fe (Zhang et al. 2016). Their $\Delta m_{15}(B)_{true}$, spectral parameters such as Si~II velocity, velocity gradient, and spectral subtypes are listed in Table 2. All spectra of the comparison are de-redshifted and corrected for the reddening from both the Milky Way and the host galaxy using the standard reddening law of Cardelli, Clayton, \& Mathis (1989), i.e., R$_{V}$ = 3.1, unless another value is reported in the previous work. For SN 2014J, an extinction of $A_\textrm{V} =1.9\pm0.6$ mag derived in \S 3.4, and R$_{V}$ = 1.6 that is averaged from different estimates (i.e., Amanullah et al. 2014, Foley et al. 2014, and Gao et al. 2015) are used to deredden its spectra. The host-galaxy reddening for the comparison SNe Ia is taken from the corresponding references for each source, which was usually derived by fitting the $B - V$ color evolution over the phases from t = +30 to 90 days past peak ($Lira-Phillips$ relation, Phillips et al. 1999) or the relation between peak $B - V$ color and decline rate $\Delta m_{15}(B)$ (i.e., Phillips et al. 1999, Wang et al. 2009b).

Spectroscopic comparisons at $\sim$$-$10 days and $\sim$$-$6 days are shown in Fig.2(a). At this earlier phase, SN 2014J is found to be similar to SN 2007le in regarding to the line profile of some main spectral features. In particular, the substructures seen in the absorption features near 4300 \AA\ and 4800 \AA\ appear quite similar in spectra of these two SNe Ia. For example, the relative strength of the double-valley absorption at $\sim$4300 \AA\, likely due to a blend of Fe {\sc ii}, Fe {\sc iii}, Mg {\sc ii} and Si {\sc iii} lines (see also Figure 12 later on in \S 4.1), is similar for SN 2014J and SN 2007le which both show stronger absorption on the blue side. The two components of such an absorption feature are comparable in strength for SN 2003du and SN 2011fe, while they seem to merge into a single component in SN 2002bo. SN 2014J and SN 2007le also exhibit similar double-valley absorption structure near $\sim$4800\AA\, with the blue-side component being relatively weak. The $\sim$4800\AA\ absorption feature appears weak and shows more substructures in SN 2003du and SN 2011fe, while it is dominated by two close absorptions with comparable strength in SN 2002bo. The spectral features from 4000 \AA\ to 6000 \AA\ will be further discussed in \S 4, where one can see that the S~{\sc ii} lines (at $\lambda$5468 and $\lambda\lambda$5612, 5654) provides a new spectral indicator to diagnose the observed diversity of SNe Ia (see detailed comparison shown in Figures 11 and 13 in \S 4.2).

The Si~{\sc ii} $\lambda$6355 absorption is one of the main spectral features in the early spectra of SNe Ia, and it appears to be asymmetric in SN 2014J, SN 2009ig, SN 2007le, and SN 2005cf at t$\sim-$10 days, suggesting the existence of the high-velocity features (HVFs). The line profiles of Si {\sc ii} become very similar by t$\sim$$-$6 days. In comparison, the Ca {\sc ii} near-infrared (NIR) triplet is the spectral feature showing the most striking differences. HVFs can be detected in the Ca {\sc ii} NIR triplet in all SNe of our sample. Among them, SN 2014J and SN 2007le show stronger HVFs of Ca II NIR triplet, while SN 2011fe shows the weakest Ca {\sc ii} HVFs. By t$\sim-$6 days, the Ca {\sc ii} HVFs weaken and the photospheric component rises in strength. Thus, we conclude that SN 2014J shows close resemblances to SN 2007le in most of the spectral features at early phases, including the Si {\sc ii}, Ca {\sc ii} lines, and absorption features at 4000-6000 \AA.

Fig.2 (b) shows the comparison near the maximum light. The difference in the 4000-6000 \AA\ region becomes smaller but the Normal/CN SNe Ia like SNe 2003du, 2005cf, and 2011fe still show substructures (or less blending) in the 4800 \AA\ absorption trough. In comparison, the Ca {\sc ii} NIR triplet still shows large diversities. For example, the Ca~{\sc ii} HVFs are still strong in SN 2005cf and SN 2014J but are almost invisible in SN 2011fe. It is not known for SN 2007co due to the short wavelength coverage. By one week after the maximum light, the spectra of different SNe Ia become relatively uniform, including the Ca {\sc ii} NIR triplet which shows large diversity at early times, but is now dominated by the more uniform photospheric components (see Fig.2(c)). The absorption features near $\sim$ 4300 \AA\ and 4800 \AA\ can still distinguish SN 2014J from the Normal/CN SNe Ia of our sample even at this phase. It is evident that SN 2014J shows fewer substructures in the above two wavelength regions, likely due to line blending, as similarly seen in the subclass of HV/BL SNe Ia like SN 2007le, SN 2007co, and 2009ig.

\subsection{Velocity and Velocity Gradient of Si II Absorption Features}
For SMilky WayN 2014J, the expansion velocity measured from the Si~{\sc ii} $\lambda$ 6355 absorption in our near-maximum-light spectrum is $v_0 = 1.22\pm0.01\times10^4$ km~s$^{-1}$. A slightly lower velocity is reported for SN 2014J from previous studies, e.g., 11,800 km s$^{-1}$ by Ashall et al. (2014) and Galbany et al. (2016), 11,900 km s$^{-1}$ by Marion et al. (2015), and 12,000 km s$^{-1}$ by Srivastav et al. (2016). All these velocities are just around the critical threshold value to discriminate the HV and normal-velocity (NV) subclasses, i.e., 11,800 km s$^{-1}$ as suggested by Wang et al. (2009a). Despite the velocity of SN 2014J locates near boundary of HV and NV subclasses, it is more reasonable to classify SN 2014J as a HV (or BL) object considering that it shows a great spectroscopic resemblance to SN 2007le.

The velocity gradient is another important parameter to describe the diversity of SNe Ia (Benetti et al. 2005). For SN 2014J, the velocity gradient is $\dot{v}=41\pm2$ km s$^{-1}$ day$^{-1}$ during the phase from t$\sim$0 to t$\sim$+10 days after the peak, which is close to the results from previous analysis (i.e., 42 km s$^{-1}$ day$^{-1}$ in Marion et al. 2015; 50 km s$^{-1}$ day$^{-1}$ in Srivastav et al. 2016; 58.4$\pm$7.3 km s$^{-1}$ day$^{-1}$ in Galbany et al. 2016). This suggests that SN 2014J can be assigned to the low-velocity gradient (LVG) subgroup in the classification scheme of Benetti et al. (2005). We further examined the evolution of line-strength
ratio of Si~II $\lambda\lambda$5958, 5979 to Si~II $\lambda$6355, defined as $R$(Si~II) (Nugent et al. 1995), and found that SN 2014J has a value ranging from $\sim$0.3 to 0.2 during the phase from t$\sim$$-$10.4 day until maximum light. This indicates that SN 2014J may be intermediate between the LVG and HVG groups (Benetti et al. 2005), while this subclassification might still suffer an uncertainty due to the lack of early-time spectra which are more sensitive to differences in explosion.

By exploring previous measurements of $v_0$ and $\dot{v}$, we find that there are only a few SNe Ia that can be put into both HV and LVG subgroups. These are SN 2002cd (with $v_{0}$(Si)=1.53$\times10^4$ km s$^{-1}$, $\dot{v}$ = 45 km s$^{-1}$ day$^{-1}$; this paper), SN 2007bd ($v_{0}$(Si)$\sim$1.26$\times 10^4$ km s$^{-1}$, $\dot{v}$=66$\pm$17 km s$^{-1}$ day$^{-1}$; Blondin et al. 2012, hereafter B12)\footnote{Blondin et al. (2012) suggested $\Delta v / \Delta t_{[+0,+10]}$ and $dv/dt$ as two different definitions of velocity gradient, and we only use their $\Delta v / \Delta t_{[+0,+10]}$ results in this paper.}, SN 2009ig ($v_{0}$(Si)$\sim$1.34$\times10^4$ km s$^{-1}$, $\dot{v}$$\sim$40 km s$^{-1}$ day$^{-1}$; Marion et al. 2013), and SN 2012fr ($v_{0}$(Si)$\sim$1.20$\times 10^4$ km s$^{-1}$, $\dot{v}$$\sim$0.3 km s$^{-1}$ day$^{-1}$; Childress et al. 2013, Zhang et al. 2014). In most cases, the HV SNe Ia also belong to the HVG subgroup and the NV ones can be generally put in the LVG subgroup (i.e., Wang et al. 2009a; Silverman et al. 2012a).

The velocity evolution of the Si {\sc ii}$\lambda$6355 absorption of SN 2014J is shown in Figure 3. Overplotted are those of some well-observed SNe Ia, including SNe 2002bo (Benetti et al. 2004), 2002dj (Pignata et al. 2008), 2003du (Stanishev et al. 2007), 2005cf (Wang et al. 2009b), 2009ig (Marion et al. 2013), 2011fe (Pereira et al. 2013; Zhang et al. 2016), and 2012fr (Childress et al. 2013; Zhang et al. 2014). The velocity evolution of SN 2002cd, SN 2007bd, SN 2007co, and SN 2007le is also shown because they are similar to SN 2014J in line profiles or velocity evolution. The mean velocity evolution of normal SNe Ia and SN 1991T-like SNe Ia obtained in Wang et al. (2009a) is also overplotted for comparison.

Although SN 2014J shows a velocity evolution that is quite similar to SN 2007le before the maximum light, it exhibits an apparently slower velocity evolution than the latter after the peak (i.e., 99 km s$^{-1}$ day$^{-1}$; B12). The velocity evolution of SN 2014J is not as slow as some extreme HV+LVG examples such as SN 2009ig and SN 2012fr. SN 2002cd is another HV+LVG example, which shows a rather large photospheric velocity. Like SN 2014J, SN 2007bd also has a moderately large $v_0$ and a relatively low $\dot{v}$, and these two SNe also have a similar velocity evolution. Although classified as a HV+LVG SN Ia, SN 2014J is still within the scope of relatively normal SNe Ia, and not as odd as those extreme events such as SNe 2002cd, 2009ig, and 2012fr. It likely constitutes a transitional SN Ia in both Benetti's and Wang's classification schemes, as mentioned in Galbany et al. (2016). SN 2007bd is another example of such a transitional SN Ia, while its spectra are less similar to SN 2014J than those of SN 2007co and SN 2007le (see \S 3.1).

In Figure 4 we compare spectroscopic and photometric parameters measured near maximum light for SN 2014J and SNe from B12. We follow the classification scheme of Branch et al.(2009), which divides SNe Ia into four groups: cool (CL), shallow silicon (SS), core normal (CN), and broad line (BL). We also divide the Branch-normal SNe Ia into NV and HV subtypes according to Wang's classification scheme. There are some overlaps between Branch's and Wang's classifications. In the CN subgroup, most objects can be put into the NV subgroup, while in the BL subgroup, 2/3 can be put into the HV subgroup and 1/3 into the NV subgroup. We notice that the SNe which belong to both the BL and NV subgroups all have a pEW of Si {\sc ii} $\lambda$6355 absorption $<$ $\sim$130 \AA, while those SNe that belong to both the BL and HV subgroups have pEW values ranging from 100 \AA\ to 200 \AA. BL+HV SNe Ia generally have a slightly weaker Si II $\lambda$5972 and stronger Si II $\lambda$6355 around the maximum light, which indicates smaller $R$(Si) (or higher photospheric temperature, Nugent et al. 1995) for these objects. CN and BL SNe (both HV and NV) have similar scatter in the pEW-$\Delta m_{15}(B)$ space. One can see that SN 2014J falls at the boundary of the CN and BL subclasses in both panels of Figure 4, but it seems to show more resemblances to BL/HV SNe Ia in terms of the substructures seen near 4000-6000 \AA, as mentioned in \S 3.1 and discussed in \S 4.1.

Considering Benetti's classification scheme, we compare spectral parameters of SN 2014J such as $v_0$ and the velocity gradient $\dot{v}$ with some photometric parameters such as $\Delta m_{15}(B)$ and $B_{\textrm{max}}-V_{\textrm{max}}$ in Figure 5. The data plotted for comparison are taken from literature \footnote{Most objects, B12; SN 2009ig, Marion et al. (2013); SN 2010ev, Guti{\'e}rrez et al. (2016); SN 2011ao \& SN 2011by, Song et al in preparation; SN 2012cg, Silverman et al. (2012b); SN 2012fr, Childress et al. (2013); SN 2013dy, Zhai et al. (2016); iPTF13ebh, Srivastav et al. (2017); SN 2015F, Cartier et al. (2017); SN 2015bp, Srivastav et al. (2017); SN 2016coj, Zheng et al. (2017)}. SNe Ia of the NV, HV, 91T-like, and 91bg-like subgroups in Wang's classification scheme are represented by different symbols. In each panel, black dashed lines mark the boundary between the LVG and HVG subgroups in Benetti's classification scheme (i.e., 70 km s$^{-1}$ day$^{-1}$) and between the NV and the HV subgroups in Wang's classification scheme (i.e. $\sim$1.2 $\times 10^4$ km s$^{-1}$).

Fig. 5(a) shows the distribution of different subclasses of SNe Ia in the $\Delta m_{15}$(B) - $\dot{v}$ space. The position of SN 2014J is closer to that of the NV subgroup, while HV SNe Ia seem to have a larger scatter in their velocity evolution, ranging from $\sim$35 km s$^{-1}$ day$^{-1}$ to about 300 km s$^{-1}$ day$^{-1}$. Benetti et al. (2005) suggested a positive relationship might exist between $\dot{v}$ and $\Delta m_{15}(B)$ in FAINT and LVG SNe Ia, which is that SNe Ia with larger $\dot{v}$ tend to have larger $\Delta m_{15}(B)$ if some events with extremely large velocity gradients ($\dot{v}>150$ km s$^{-1}$ day$^{-1}$) such as SNe 2001ay, 2002bf, 2004dt, and 2006X are excluded. We fit a linear relationship as $\Delta m_{15}(B)$=0.00340($\pm$0.00072)$\times\dot{v}$-0.978($\pm$0.053) mag, with a Pearson correlation coefficient $r=0.53$. Fig. 5(b) shows the velocity gradient and velocity distribution of our sample. Wang et al. (2009a) suggested that the velocity gradient $\dot{v}$ is positively correlated with the near-maximum-light velocity $v_0$, and this correlation was later confirmed by Silverman et al. (2012a). Notice that there are several outliers that do not follow such a trend, including SN 2009ig and SN 2002cd, which both have large $v_0$ but smaller$ \dot{v}$. Objects showing such characteristics can be subclassified as HV+LVG SNe Ia, and SN 2014J might be also a member of this subgroup.

In Fig. 5(c) and Fig. 5(d), we examined the correlations between the $B - V$ colour at maximum light with $\dot{v}$ and $v_0$, respectively. The colours are corrected only for the reddening in the Milky Way. SN 2014J shows a very red colour, due to the large extinction in M82, which will be further discussed in \S 3.4. After excluding those SNe that are very likely suffering significant reddening in their host galaxies (i.e., $B_{\textrm{max}}-V_{\textrm{max}}>0.2$, as marked by the black dashed lines), we conducted linear fits to the observed relations between the peak $B - V$ colours, $\dot{v}$, and $v_0$ for the rest sample, as indicated by the cyan lines. For the $(B_{\textrm{max}}-V_{\textrm{max}})$ vs $\dot{v}$ correlation, we derive $B_{\textrm{max}}-V_{\textrm{max}}$=0.015($\pm$0.019)$\times [\dot{v}$/100 km s$^{-1}$ day$^{-1}$]+0.004($\pm$0.018) mag, with a Pearson correlation coefficient $r=0.12$. For the $(B_{\textrm{max}}-V_{\textrm{max}})$ vs $v_0$ correlation, we derive $B_{\textrm{max}}-V_{\textrm{max}}$=0.31($\pm$0.06$)\times [v_0/10^4$ km s$^{-1}]-$0.33($\pm$0.07) mag, with the Pearson coefficient being 0.33. The tendency that SNe Ia with larger $v_0$ have redder $B_{\textrm{max}}-V_{\textrm{max}}$ colour (see Fig. 5(d)) was initially reported by Wang et al. (2009a) and later confirmed by B12\footnote{Note that these relationship and hypothesis are not valid for the SNe with large $B - V$ colours.}. The weaker correlation of $B_{\textrm{max}}-V_{\textrm{max}}$ with $\dot{v}$ may be related to small number statistics and/or that the physics affecting $\dot{v}$ is more complex than that for $v_0$. Benetti et al. (2005) also suggested that the HVG SNe Ia may have more efficient mixing of heavy elements and hence larger opacities in the outer layers, which can cause lower temperatures for the photosphere. Different photospheric temperatures imply different intrinsic colours for SNe Ia. On the other hand, Wang et al. (2009a) proposed that the redder $B - V$ colours revealed for SNe Ia with large Si {\sc ii} velocities are likely due to that HV SNe Ia suffered larger extinction from their host galaxies than their NV counterparts. This is consistent with the observation that the SNe Ia showing systematically blueshifted velocity structures (or outflow) due to the circumstellar materials tend to be the HV subclass (Sternberg et al. 2011).

\subsection{Late-Time Spectra}

Four late-time spectra of SN 2014J at $t=+296.3$ d, $+404.2$ d, $+428.2$ d, and $+473.2$ d after the maximum light are plotted in Figure 6. We also compare these spectra with those of some normal SNe Ia, such as SNe 1998bu (Cappellaro et al. 2001), 2002dj (Pignata et al. 2008), 2003du (Stanishev et al. 2007), 2006X (Wang et al. 2008), and 2011fe (Zhang et al. 2016). Forbidden lines of iron-group elements (IGEs) dominate the emission features in the nebular spectra of these SNe Ia. The prominent emission features near 4400 \AA, 4700 \AA, and 5200 \AA\ are due to [Fe {\sc ii}] and [Fe {\sc iii}] lines. The minor peak at $\sim$ 5900 \AA\ is formed by a [Co {\sc iii}] line. Two emission features at $\sim$ 7200 \AA\ and 7500 \AA\ with comparable strength are formed by [Fe {\sc ii}] and [Ni {\sc ii}] lines. The main emission features in the nebular spectra of these SNe Ia are quite similar. It can be seen that the late-time spectra of SN 2014J show relatively stronger emission in the wavelength region from 4000-5000 \AA, resembling that of SNe Ia affected by dust scattering, such as SN 1998bu (Cappellaro et al. 2001) and SN 2006X (Wang et al. 2008). The detection of a light echo from SN 2014J has been reported by Crotts (2015) and Yang et al. (2017) using the late-time images taken with the Hubble Space Telescope at a few hundred days after the explosion of this SN. However, the late-time spectra can not ensure a detection of the light echo in SN 2014J due to the uncertainty in reddening correction and possible contamination from the host galaxy, our forthcoming paper II (Li et al. in preparation) will focus on the studies of the late-time photometry of SN 2014J.

The nebular spectra can be used to study the inner region of SNe Ia. The FWHM of Fe emission line at $\sim$ 4700 \AA\ serves as an indicator of the kinetic energy of the explosion. For example, Mazzali et al. (1998) found a relation between the FWHM of [Fe {\sc iii}] $\lambda$4700 and $\Delta m_{15}(B)$, while B12 suggested that there is no such a relation. By fitting a Gaussian above the continuum suggested in Mazzali et al (1998), we measured the FWHM of [Fe {\sc iii}] as 12.8$\pm$1.6$\times$10$^3$km s$^{-1}$ in the GTC spectrum taken at $t=+428.2$ d of SN 2014J. This value is among the highest values listed in Table 7 of B12, suggesting that SN 2014J may have a larger kinetic energy relative to other SNe Ia. The velocity shifts of forbidden lines [Fe {\sc ii}] $\lambda$7155 and [Ni {\sc ii}] $\lambda$7378 have been suggested as an evidence of off-centre explosion by Maeda et al. (2010). For SN 2014J, we derive $v_{\textrm{neb}}$([Fe {\sc ii}] $\lambda$7155)=$2.2\times 10^3$ km s$^{-1}$ and $v_{\textrm{neb}}$([Ni {\sc ii}] $\lambda$ 7378)=$2.8\times 10^3$ km s$^{-1}$ from the $t\sim428$ day spectrum, by using a 2-Gaussian function to fit the double emission lines superimposed on the continuum. The redshift of these lines suggests that SN 2014J is viewed from the opposite direction of the ignition side, following the trend seen in other HV SNe Ia (Maeda et al. 2010). However, whether the measured velocity shift really indicates an asymmetric explosion is still controversial (e.g., Wang et al. 2013).

\subsection{High-Resolution Spectrum and Extinction}
In this section, we will discuss the Na {\sc i} D, K {\sc i}, and DIB features in the high-resolution spectrum of SN 2014J. Note that the Na {\sc i} doublet absorption and some diffuse DIB absorption features (i.e., at 5780, 5797 and 6613 \AA) can be also detected in our intermediate-resolution spectra, as shown in Figure 7. In particular, the two components of Na {\sc i} D can be clearly resolved and they did not show any significant variations, as indicated by the results listed in Table 3. On the other hand, it is hardly to tell from the intermediate-resolution spectra whether the DIB features evolve or not due to the lower S/N ratio and uncertainties in wavelength calibration of the spectra. We thus focus on the high-resolution spectrum in the following analysis.

Our high-resolution spectrum was taken on 2014 Jan 29, about 3 days before maximum light. Its wavelength range is from $\sim$ 4347 \AA\ to 9175 \AA. Owing to the deficiency of the coverage at short wavelengths, we can not study the spectral features of Ca {\sc ii} H\&K, CN, CH, CH$^+$. The Li {\sc i} absorption cannot be studied because it falls in a gap between different spectral orders. Figure 8 shows four DIBs detected at 5780, 5797, 6283, and 6613 \AA\ in our spectrum. All these features have been first normalized to the continuum using low-order polynomials. The full widths at half maximum (FWHMs) and $EW$s calculated from the Gaussian fit are listed in Table 4, which are consistent with the results reported in G15\footnote{Note that the FWHMs shown in G15 should be multiplied by a factor of 2.35, after private communication with Graham in 2017.}

Figure 9 shows the Na {\sc i} doublet absorptions from our high-resolution spectrum and the spectra obtained by G15 on Jan. 22, Jan. 27, and Feb. 24, 2014, respectively. There are two main absorption features due to the Milky Way, with velocities of $\sim$ $-$50 km s$^{-1}$ and $\sim$ 0 km s$^{-1}$, respectively, in the observed frame. Several absorption features from M82 have velocities ranging from 0 to $-$150 km s$^{-1}$ in the restframe of M82. G15 showed that the Na {\sc i} D absorption features did not show any significant evolution with time. From the analysis of our spectrum, we agree with their conclusions.

The K {\sc i} $\lambda\lambda$7664.90, 7698.96 doublet absorption lines can also be detected in our high-resolution spectrum, as shown in Figure 10. The absorption features of the K {\sc i} $\lambda\lambda$7664.90, 7698.96 doublet lines have many components with velocities ranging form 0 to $\sim$ $-$150 km s$^{-1}$, with the strength of K {\sc i} $\lambda$7698.96 lines being lower than that of K {\sc i} $\lambda$7664.90 line. The $\lambda$7698.96 absorption in the high-resolution spectrum from G15 (taken on Jan 27, 2014) is overplotted for comparison, which matches the main features in our spectrum but with stronger absorptions. This change within 2 days is more likely due to the poor $S/N$ ratio of our spectrum.

Since the DIB absorptions are likely caused by the ISM molecules (e.g., Herbig 1995 for a review), they have also been regarded as better indicators of ISM extinction (i.e., Friedman et al. 2011, Phillips et al. 2013, Welty et al. 2014). Based on the DIB features detected in the high-resolution spectra of SN 2014J, Welty et al. (2014) estimated the visual extinction as $A_\textrm{v}$ $\sim$ 1.9$\pm$0.2 mag, while G15 found $A_\textrm{v}^{\textrm{host}}$ =1.8$\pm$0.9 mag. Goobar et al. (2014) deduced an extinction value of $A_\textrm{v}^{\textrm{host}}$=2.5$\pm$1.3 mag (i.e., Phillips et al. 2013). From the measurement of our spectrum, we derive an extinction of $A_\textrm{v}$$\sim$1.9$\pm$0.6 mag for SN 2014J, based on the average of the relations obtained by Friedman et al. (2011) and Phillips et al. (2013). We do not count in the relation obtained by Welty et al. (2014) due to its lack of the systematic uncertainty. This extinction value derived from our high-resolution spectrum is consistent with previous estimates.

\section{DISCUSSIONS}
In this section, we will further discuss the unusual spectral features seen in SN 2014J and explore their possible origin. There are numerous studies focusing on the diversity of Si~{\sc ii} and Ca~{\sc ii} features based on large samples of SNe Ia (e.g., Silverman et al. 2015; Zhao et al. 2015), and even on the O~I absorptions in the early spectra of SNe Ia (Zhao et al. 2016). However, few studies have been conducted for the spectral features in wavelength region from 4000-6000 \AA\ except for the Si~{\sc ii} $\lambda$ 5972 absorption. Figure 11 shows the comparison of the wavelength regions 4000-5200 \AA, 5000-5600 \AA, and 5500-6400 \AA, between $t\sim-$11 d and $-$8 d. The latter two wavelength regions cover the S~{\sc ii} and Si {\sc ii} absorptions, respectively. Spectra of SN 2014J are shown in black, while the spectra of HV, NV, 91T-like subtypes are represented with the red, blue, magenta lines, respectively. Green and purple lines are for SNe 2009dc (Super-Chandrasekhar SN Ia) and 2012cg (99aa-like), respectively.

\subsection{Absorption features at 4000-5200 \AA}

We first examine the spectral features at 4000-5200 \AA, where there are two primary absorptions at $\sim$ 4300 \AA\ and 4800 \AA\, respectively. Owing to different velocities, we are not sure whether these features at similar positions in the spectra in different SNe are produced by the same elements or not. In order to further study these features, we fit two spectra of SN 2014J in the range of 4000-5200 \AA\ at $-$10.4 d and +0.6 d using \texttt{SYNAPPS} (Thomas et al. 2011). We adopt ions of both intermediate mass elements (IMEs), such as Mg {\sc ii}, Si {\sc ii}, Si {\sc iii}, S {\sc ii}, and iron group elements (IGEs) such as Fe {\sc ii}, Fe {\sc iii}, Co {\sc ii}, Ni {\sc ii}. The best-fit model spectra are shown in Figure 12. Overplotted are the spectra of SNe 2002bo, 2007co, and 2011fe. The photometric temperatures and velocities given by \texttt{SYNAPPS} are $T_{phot}=11.2\times 10^3$ K, $v_{phot}=14.0\times 10^3$ km s$^{-1}$ for $t=-$10.4 d and $T_{phot}=10.8\times 10^3$ K, $v_{phot}=11.6\times 10^3$ km s$^{-1}$ for $t=$+0.6 d.

At $t\sim -$10 days, the absorption at $\sim$4260 \AA\ can be due to Mg~{\sc ii} and Fe~{\sc iii} lines, while the Si~{\sc iii} line can contribute to the minor absorption at $\sim$4380 \AA. The absorption feature near $\sim$4800 \AA\ can be due to blending of Ni {\sc ii}, Si~{\sc ii}, S~{\sc ii}, and Fe {\sc iii} lines. Ni {\sc ii} is likely responsible for the strong absorption at $\sim$4800 \AA, while Fe {\sc iii} accounts for the absorption at $\sim$4950 \AA. At $t\sim$0 days, stronger Mg {\sc ii} absorption dominates the feature at $\sim$4260 \AA, with only a small contribution from the Fe {\sc iii} line. The Fe {\sc iii} absorption at $\sim$4950 \AA\ also becomes weak at this phase when the Si {\sc ii}, Fe {\sc ii}, Ni {\sc ii} lines are the main contributor to the broad absorption through at $\sim$4600-5100 \AA. The weakening of Fe {\sc iii} and the strengthening of Fe {\sc ii} are consistent with the decrease of the photospheric temperature with time.

With the above understanding of possible ion contributions to the spectral features at 4000-5200 \AA, we then examine the diversity seen in different subtypes of SNe Ia including the transitional object SN 2014J. We first notice that the 4300 \AA\ absorption feature seen in the $t\sim -$10 day spectrum of SN 2002bo has only one major trough at $\sim$4260 \AA, formed by Mg~{\sc ii} and Fe~{\sc iii} lines. In comparison, other HV SNe Ia, such as SN 2007le, SN 2009ig, and SN 2014J, have two troughs near 4300 \AA\ at this phase, with the major trough from Mg {\sc ii} and Fe {\sc iii} and the minor one likely formed by Si {\sc iii} (see Figure 12). One can see that these two absorption components near 4300 \AA\ have large variation for the HV subclass at early phases, while they show little difference in NV SNe Ia. On the other hand, SN 2003du, SN 2009ig, SN 2009dc, SN 2012fr, and perhaps SN 2012cg seem to have stronger Si {\sc iii} absorption troughs (near 4400 \AA) at similar phases, and all of them are characterized by a slow velocity evolution after the maximum light (see also Figure 13). In particular, SN 2009ig also belongs to the HV subclass and has an ejecta velocity similar to SN 2002bo, but much weaker Si {\sc ii} $\lambda$5972 as shown in the right panel of Figure 11.

The absorption trough near 4800 \AA\ also shows some diversity at $t\sim -$10 days. This feature is very weak in SN 2009dc, SN 2009ig, SN 2012cg, and SN 2012fr, but is very strong in SN 2002bo. Note that the main part of the 4800\AA\ absorption feature is absent in SN 1991T which shows a strong absorption at $\sim$4900 \AA\ (due to prominent Fe {\sc iii} lines). The formation of the 4800 \AA\ absorption feature is complex, due to blending of many lines such as Si {\sc ii}, S {\sc ii}, Fe {\sc ii}, Fe {\sc iii}, and Ni {\sc ii}, as indicated by the \texttt{SYNAPPS} fitting as shown in Figure 12. For SN 2014J, the absorption trough can be decomposed by a minor absorption due to Si {\sc ii} at $\sim$4690 \AA\ and a major absorption from blending of S {\sc ii}, Si {\sc ii}, and Ni {\sc ii} lines at $\sim$4820 \AA. For SN 2002bo, the strong Si {\sc ii} absorption seen at $\sim$4700 \AA\ is consistent with the weak Si III absorption at $\sim$4400 \AA, suggesting a lower temperature. We notice that SN 2005cf has a very strong HVF of Si {\sc ii} $\lambda$ 6355 while from 4000 \AA\ to 5200 \AA\ it shows very similar features as SN 2011fe, which shows no HVF of Si {\sc ii} $\lambda$6355. This indicates that the features at 4000 \AA\ to 5200 \AA\ are possibly unrelated to the outmost ejecta.

\subsection{S~{\sc ii} "W"-Shaped Feature}
The middle panel of Figure 11 shows the comparison of the wavelength region covering the S~{\sc ii} "W" absorptions (5000-5600 \AA). We notice that the blue component, S {\sc ii} $\lambda$5468, is much weaker than the red component, S {\sc ii} $\lambda$5640 for most SNe. In some SNe Ia such as SN 2002bo, SN 2003W, and SN 2011fe, the S {\sc ii} $\lambda$5640 absorption seems to be relatively strong, while the S {\sc ii} $\lambda$5468 absorption are relatively prominent in the spectra of SN 2003du, SN 2009dc, SN 2012cg, SN 2012fr, and SN 2014J. For the latter sample, the two components of the S {\sc ii} "W" structure are more comparable in strength and appear to be more symmetric than the rest SNe Ia in comparison. One common feature for those SNe Ia showing relatively strong S {\sc ii} $\lambda$5468 absorption is that their Si {\sc ii} $\lambda$5972 absorptions tend to be weaker (or barely seen), as shown in the right panel of Figure 11.

To better quantify the observed diversity of the S{\sc ii} "W" lines in SNe Ia, we define the ratio of the equivalent widths (EWs) of S {\sc ii} $\lambda$5468 and S {\sc ii} $\lambda$5640 measured at $t\sim-$10 days as $R$(S). Following the method described in Zhao et al. (2015), we apply a double gaussian fit to the observed line profiles of S {\sc ii} $\lambda$5468 and S~{\sc ii}$\lambda$5640 to measure their $EWs$s. As shown in left panel of Figure 13, brighter SNe with smaller $\Delta m_{15}(B)$ generally have larger $R$(S), while fainter SNe tend to have smaller values. The Pearson correlation coefficient $r = -0.75$ suggests this anti-correlation is significant. This anti-correlation of the measured $R$(S) with $\Delta m_{15}(B)$ indicates that the observed diversity of the S~{\sc ii} "W" feature may be related to the SN luminosity.

From the right panel of Figure 13, one can see that the $R$(S) shows a weaker anti-correlation with $\dot{v}$(Si), with the Pearson correlation coefficient being as $r$=$-$0.54. The LVG SNe Ia, especially those with $\dot{v}$(Si)$\lesssim$40 km s$^{-1}$ day$^{-1}$, tend to have on average larger $R$(S). The $R$(S)- $\dot{v}$(Si) plot can help further identify different subclasses of SNe Ia under current classification schemes. For example, SN 2014J and SN 2007le are spectroscopically quite similar and can be put into the HV subclass, but the former has a smaller $\dot{v}$(Si) and larger $R$(S). SN 2002cd+SN 2003W and SN 2009ig+SN 2002bo are other two pairs of such examples. In each pair the SNe have similar Si {\sc ii} velocities and can be classified as HV SNe Ia, but the former ones (SN 2002cd and SN 2009ig) have smaller $\dot{v}$(Si) and larger $R$(S) relative to the latter ones (SN 2003W and SN 2002bo). These examples demonstrate that the Si~{\sc ii} velocity and the velocity gradient $\dot{v}$(Si) are not completely linked to each other. Given the correlation between $R$(S) and $\dot{v}$(Si), $R$(S) may be used as an additional parameter to describe the diversity of SNe Ia.

\subsection{Physical Constraints}

The relationship between $R$(S) and $\Delta m_{15}(B)$ can be interpreted as a temperature/ionization effect. A higher photospheric temperature $T_{\textrm{eff}}$ may also lead to the formation of a prominent Si~{\sc iii} absorption at $\sim$ 4400\AA. However, other mechanisms may be also needed given the fact that S~{\sc ii} $\lambda$5468 and S~{\sc ii} $\lambda$5640 have little difference in their excitation energies (i.e., 13.6 eV versus 13.7 eV). In particular, the correlation between $R$(S) and $\dot{v}$(Si) indicates that the ratio of the S~{\sc ii} lines is likely due to other mechanisms, e.g., the explosion physics and/or progenitor scenarios.

In theory, for the typical W7 or the delayed detonation model, it is hard to produce a SN Ia with a long plateau in velocity or a low velocity gradient. In contrast, a shell-like density structure can naturally produce such a slowly-evolved velocity as the photosphere can remain in the shell for some time (Khokholov et al. 1993; Hoeflich \& Khokholov 1996). Quimby et al. (2007) modeled the effects from the former two scenarios on the observed Si~{\sc ii} velocity and found that a spherically symmetric shell with a mass of 0.2 M$_{\odot}$ can reproduce the flat velocity evolution seen in SN 2005hj. Note that this SN is also found to be a normal SN Ia with a relatively strong S {\sc ii} $\lambda$5460 absorption and hence a large $R$(S). Shell masses of about 0.04, 0.1, 0.12, and 0.13 M$_{\odot}$ can be derived for SN 2002cd, SN 2009ig, SN 2012fr, and SN 2014J, respectively, according to the results from models (i.e., Quimby et al. 2007) if their slow velocity evolution is also caused by the shell-like density structure.

\section{CONCLUSIONS}

In this paper, we present extensive optical spectroscopy (including one very high-resolution spectrum) for SN 2014J, covering the phases from $-$10.4 d to +473.2 d from maximum light. Based on the detected DIBs in our high-resolution spectrum, we derive a total extinction $A_\textrm{v}$ $\sim$1.9$\pm$0.6 mag for SN 2014J.

The spectral properties of SN 2014J are overall similar to that of SN 2007co and SN 2007le. It can be classified as a HV SN Ia, but also belongs to the LVG subclass, which is somewhat against the general trend that HV SNe Ia tend to have a larger velocity gradient. Based on the spectral features in the wavelength region 4000-6000 \AA\, we suggest that SN 2014J, SN 2002cd, and SN 2009ig (and perhaps SN 2012fr) share some common properties, and may represent a new HV subtype with a flat velocity evolution (dubbed as HV+LVG subclass). We caution, however, that the classification of SN 2014J as a member of HV+LVG SNe Ia is not as robust as SN 2002cd and SN 2009ig due to that it locates near the boundary of HV and NV subgroups. It is possible that SN 2014J represents a transitional object linking the HV+LVG and NV+LVG subgroups.

The HV+LVG objects are found to have higher temperatures, as evidenced by the strong Si {\sc iii} feature at $\sim$4300 \AA\ and the weak Si {\sc ii} $\lambda$5972 absorption seen in their early spectra. Moreover, we find that the ratio of S~{\sc ii} $\lambda$5468 and S~{\sc ii} $\lambda$5640 absorption, $R$(S), measured at $t\sim-$10 days, is inversely correlated with $\Delta m_{15}(B)$. The $R$(S) parameter also shows an inverse correlation with the velocity gradient, but with large scatter at the LVG end. Additional mechanism is needed to account for these anti-correlations of $R$(S), as these two lines of S~{\sc ii} have similar excitation energies. Theoretically, the slow velocity evolution can be formed in a shell-like density structure produced by pulsating delayed detonation scenario, mergers, or interaction of ejecta with circumstellar materials. These scenarios are also consistent with the high photospheric temperatures observed in the HV+LVG subtype of SNe Ia, but a large, well-observed sample of similar properties is needed to test current models. Detailed modeling of W-shaped S~{\sc ii} lines may also help make further distinguish between different scenarios.

\section*{Acknowledgements}

We thank the anonymous referee who provided suggestive comments to further improve the paper. We acknowledge the support of the staffs of the Lijiang 2.4~m telescope and Xinglong 2.16~m telescope. This paper is also partially based on observations collected at Copernico 1.82m telescope of the INAF Osservatorio Astronomico  di Padova; Galileo 1.22m telescope of the University of Padova; the Gran Telescopio Canarias (GTC), installed in the Spanish Observatorio del Roque de los Muchachos in the island of La Palma of the Instituto de Astrofisica de Canarias, Spain; and the 3.6 m Italian Telescopio Nazionale Galileo (TNG) operated by the Fundaci\'on Galileo Galilei - INAF on the island of La Palma, Spain. Funding for the LJT has been provided by Chinese Academy of Sciences and the People's Government of Yunnan Province. The LJT is jointly operated and administrated by Yunnan Observatories and Center for Astronomical Mega-Science, CAS. This work is supported by the Major State Basic Research Development Program (2013CB834903), the National Natural Science Foundation of China (NSFC grants 11178003, 11325313, and 11633002), and the National Program on Key Research and Development Project (Grant NO. 2016YFA0400803)". K.-C. Zhang is supported by the China Scholarship Council (CSC, NO. 201706210140). T.-M. Zhang is supported by the NSFC (grants 11203034). This work was also partially Supported by the Open Project Program of the Key Laboratory of Optical Astronomy, National Astronomical Observatories, Chinese Academy of Sciences. J.-J. Zhang is supported by the NSFC (grants 11403096, 11773067), the Youth Innovation Promotion Association of the CAS (grants 2018081), the Western Light Youth Project, and the Key Research Program of the CAS (Grant NO. KJZD-EW-M06). LT, SB are supported by the PRIN-INAF 2014 with the project "Transient Universe: unveiling new types of stellar explosion with PESSTO". ST acknowledges support by TRR 33 "The Dark Universe" of the German Research Foundation.

{}

\begin{figure*}
\includegraphics[width=\textwidth]{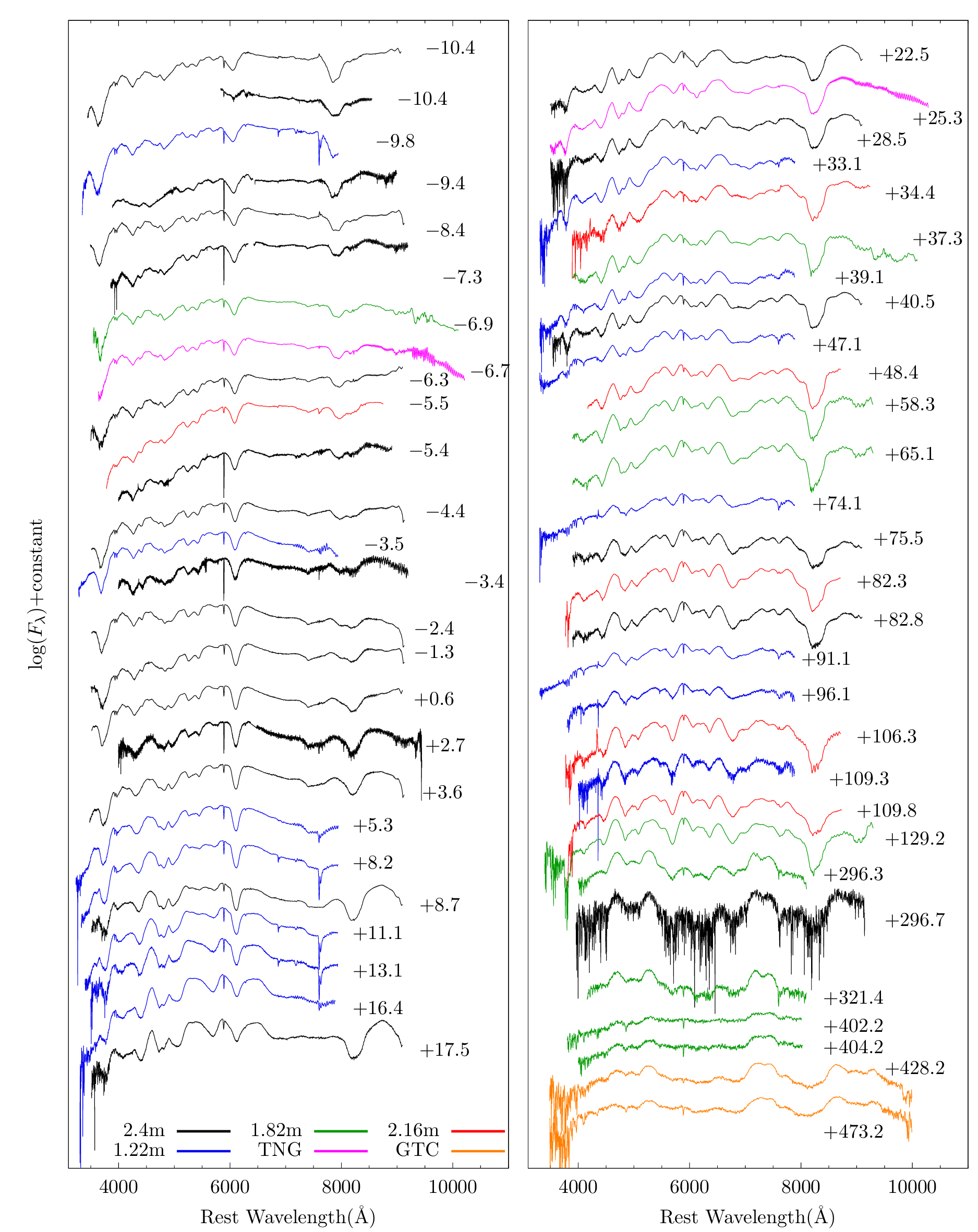}
\caption{Optical spectra of SN 2014J. Black, red, magenta, green, blue, and orange lines are spectra from Lijiang 2.4~m telescope of YNAO, Xinglong 2.16~m telescope of NAOC, the 3.58-m Telescopio Nazionale Galileo (TNG), the Copernico 1.82-m telescope of the INAF-Padova Observatory, and the Galileo 1.22-m telescope of the Padova University, and the Gran Telescopio Canarias (GTC) 10.4~m telescope, respectively.}
\end{figure*}

\begin{figure*}
\includegraphics[width=\textwidth]{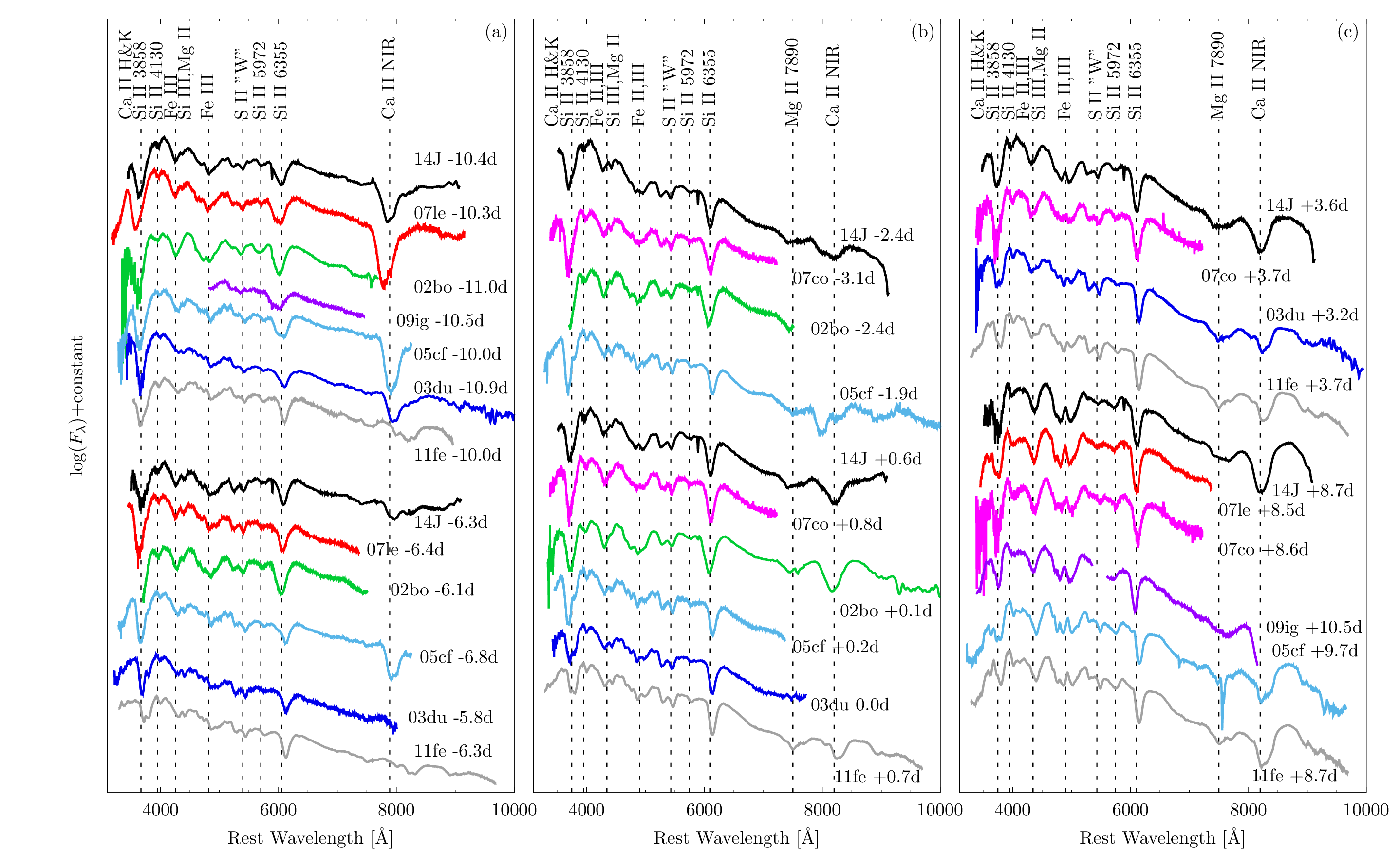}
\caption{Spectral comparison of SN 2014J with some normal SNe Ia with similar $\Delta m_{15}(B)$, such as SNe 2002bo, 2003du, 2005cf, 2007co, 2007le, 2009ig, and 2011fe, from $\sim -10$ d to $\sim +9$ d. All the spectra are dereddened. An extinction of $E(B-V)=1.23$ and $R_V=1.46$ for SN 2014J (Marion et al. 2015) is adopted.}
\end{figure*}

\center
\begin{figure*}
\includegraphics[width=0.7\textwidth]{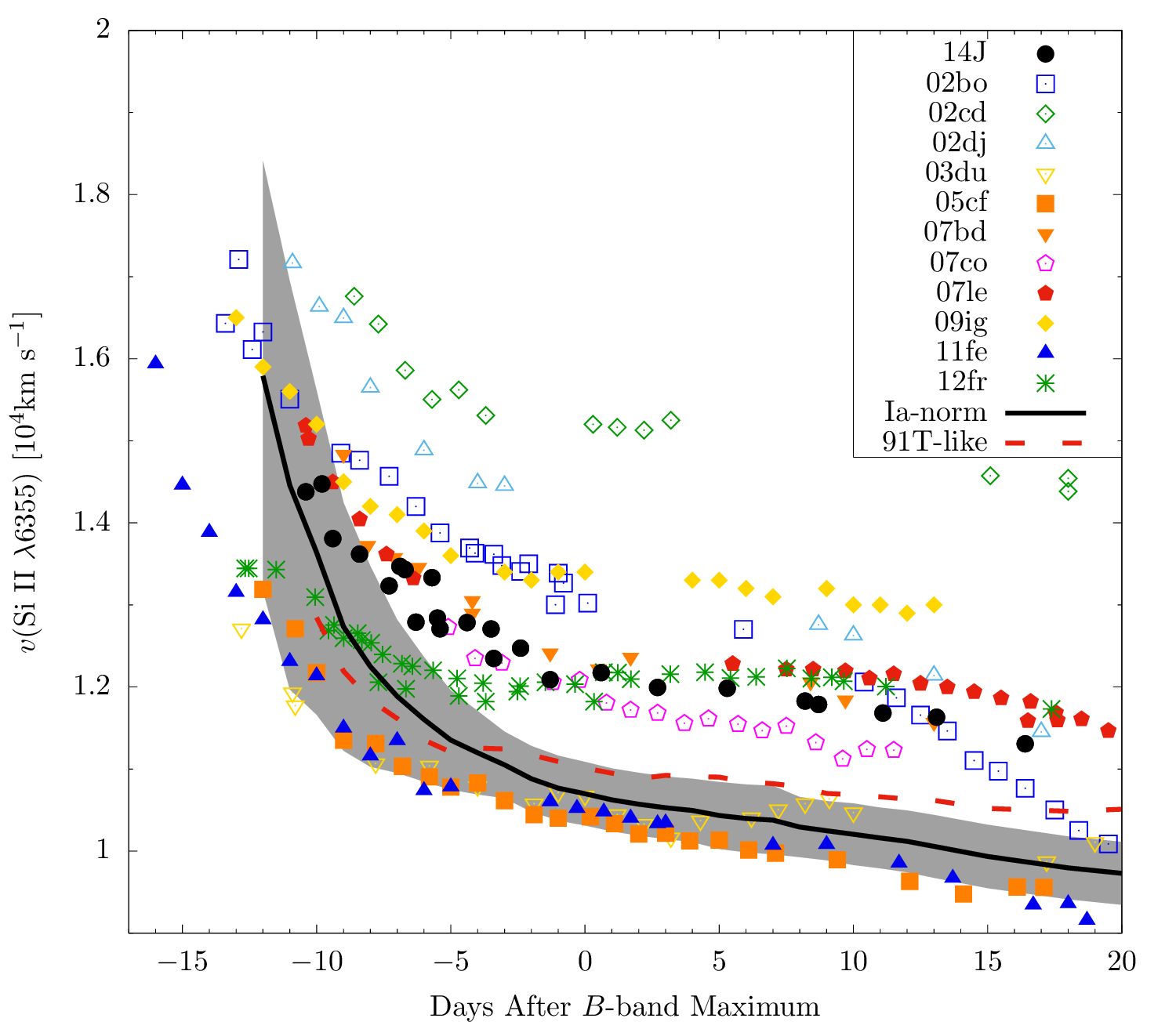}
\caption{Temporal evolution of the expansion velocity of PVFs for the Si II $\lambda$6355 absorption of SN 2014J and some other SNe Ia. The solid line shows the mean evolution of well-observed normal SNe Ia, and the gray region represents the 1$\sigma$ uncertainty; the dashed red line illustrates the evolution of the mean velocity for 1991T-like SNe. Both lines are from Figure 1 in Wang et al. (2009a).}
\end{figure*}

\clearpage
\begin{figure*}
\includegraphics[width=1.0 \textwidth]{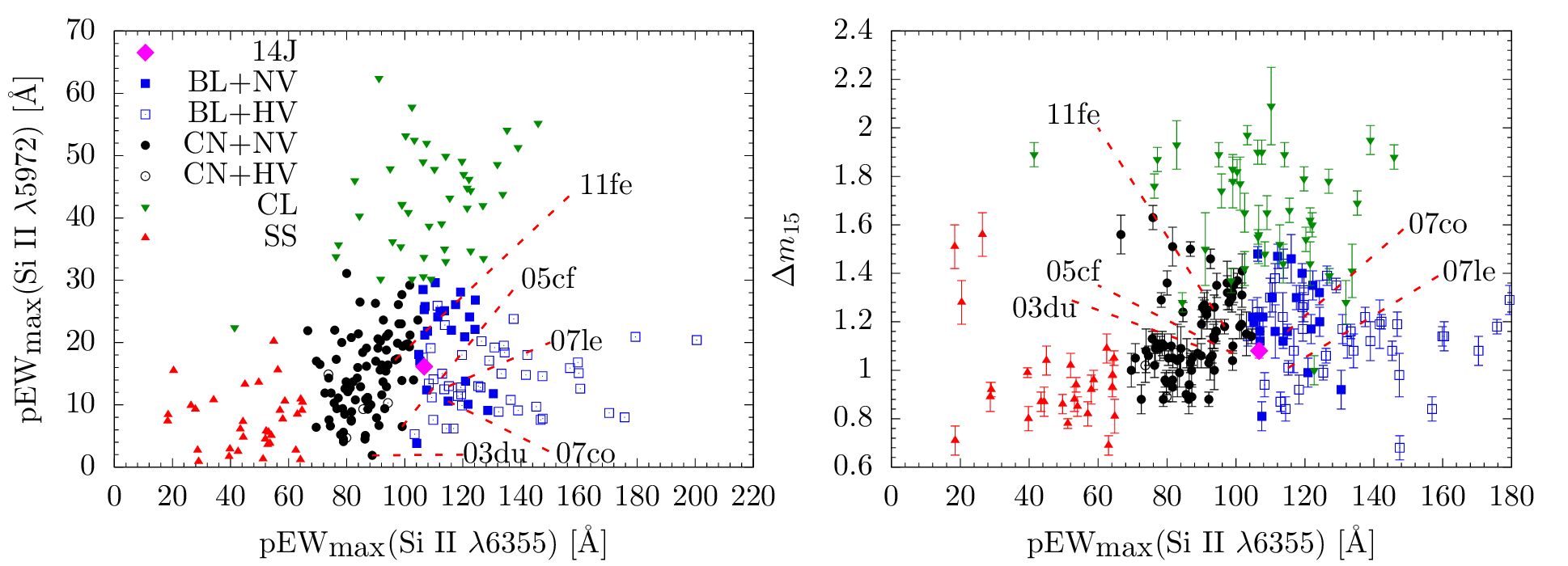}
\caption{Comparison of the EW of Si II $\lambda$5972, $\lambda$6355 and $\Delta m_{15}(B)$ of SN 2014J with other SNe Ia in the classification scheme of Branch et al. (2009) and Wang et al. (2009a). Data in this diagram are collected from Blondin et al. (2012).}
\end{figure*}

\clearpage
\begin{figure*}
\includegraphics[width=\textwidth]{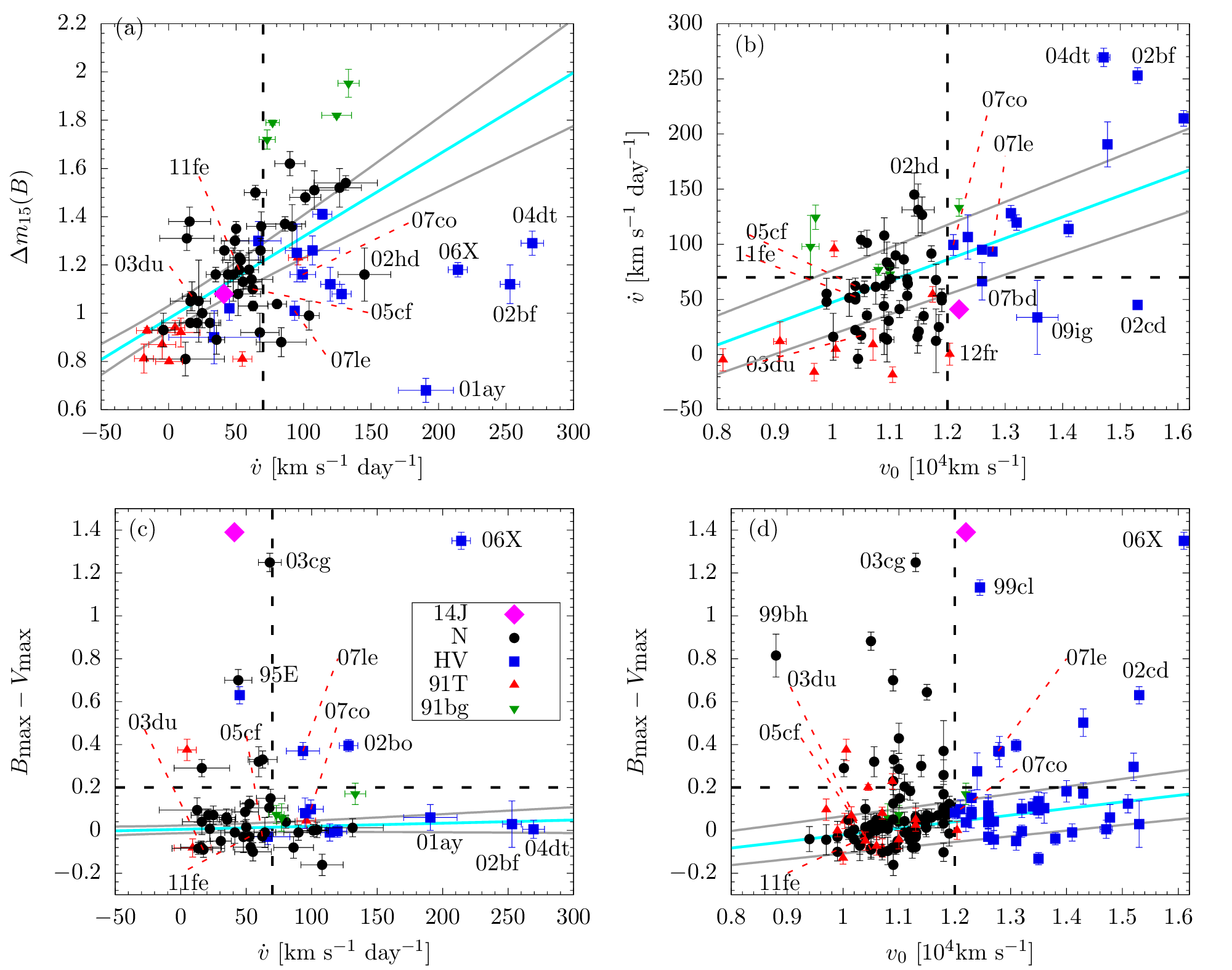}
\caption{Comparison between the spectral parameters $v_0$ and $\dot{v}$ with photometric parameters like $\Delta m_{15}(B)$ and $B_{\textrm{max}}-V_{\textrm{max}}$. SNe of the NV, HV, 91T-like, and 91bg-like subgroups in Wang's classification scheme are represented with different symbols. In each panel, black dashed lines mark the boundaries of Benetti's classification scheme (70 km s$^{-1}$ day$^{-1}$) and Wang's classification scheme (1.2$\times10^4$ km s$^{-1}$). A colour cut of $B_{\textrm{max}}-V_{\textrm{max}}=0.2$ mag is shown by the black dashed lines in panels (c) and (d). The cyan lines in each plot represent the best linear fits to the observed data. Grey lines show the 1$\sigma$ confidence intervals of these fits.}
\end{figure*}

\clearpage
\begin{figure*}
\includegraphics[width=\textwidth]{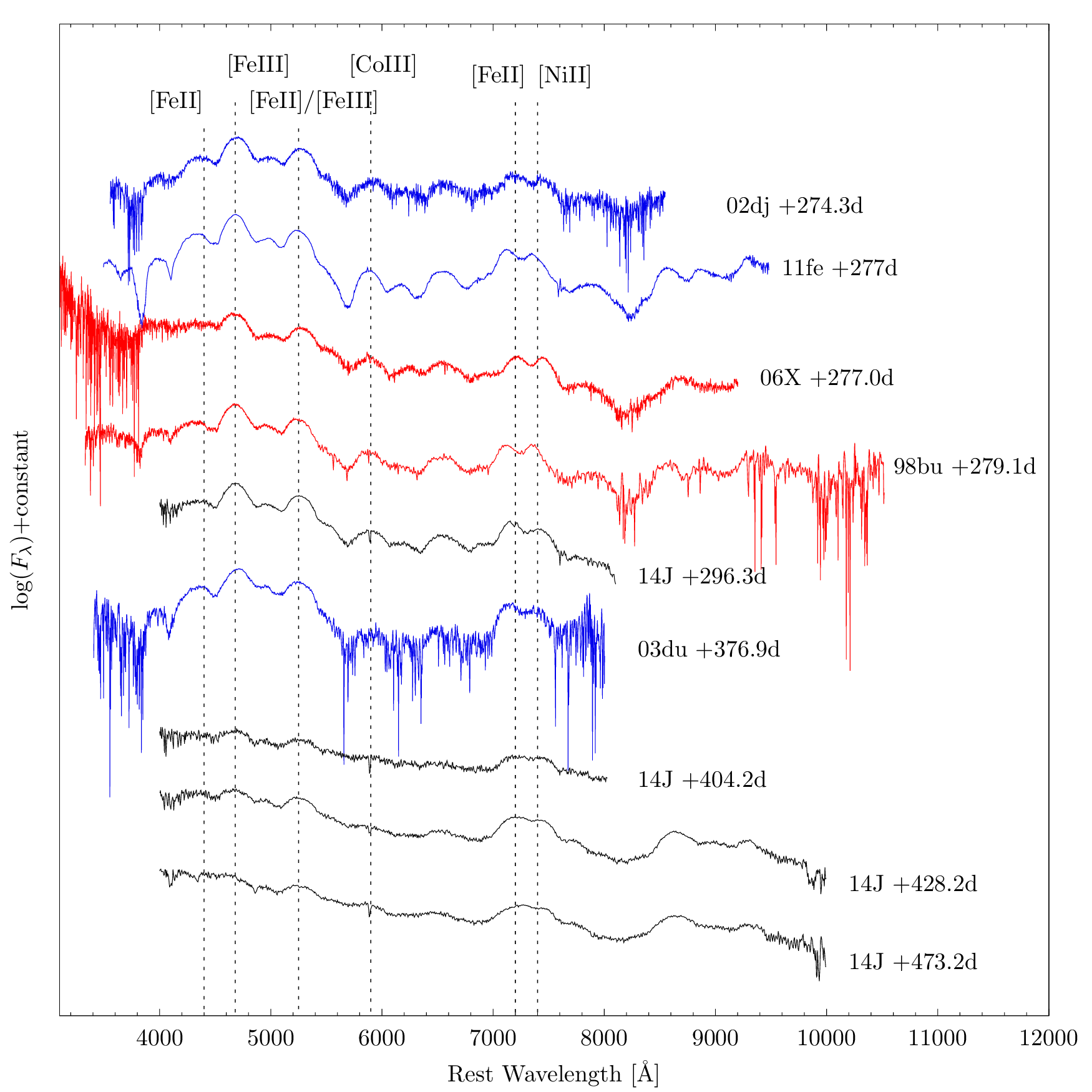}
\caption{Late-time (t $>$ +250 d) spectra of SN 2014J (in black), compared with some SNe with (red) and without (blue) light echo detections. The spectrum of SN 2014J at t=+296.7 d is binned by 30 \AA. All the spectra are properly dereddened. An extinction of $E(B-V)=1.23$ and $R_V=1.46$ for SN 2014J (Marion et al. 2015) is adopted.}
\end{figure*}

\clearpage
\begin{figure*}
\includegraphics[width=0.5\textwidth]{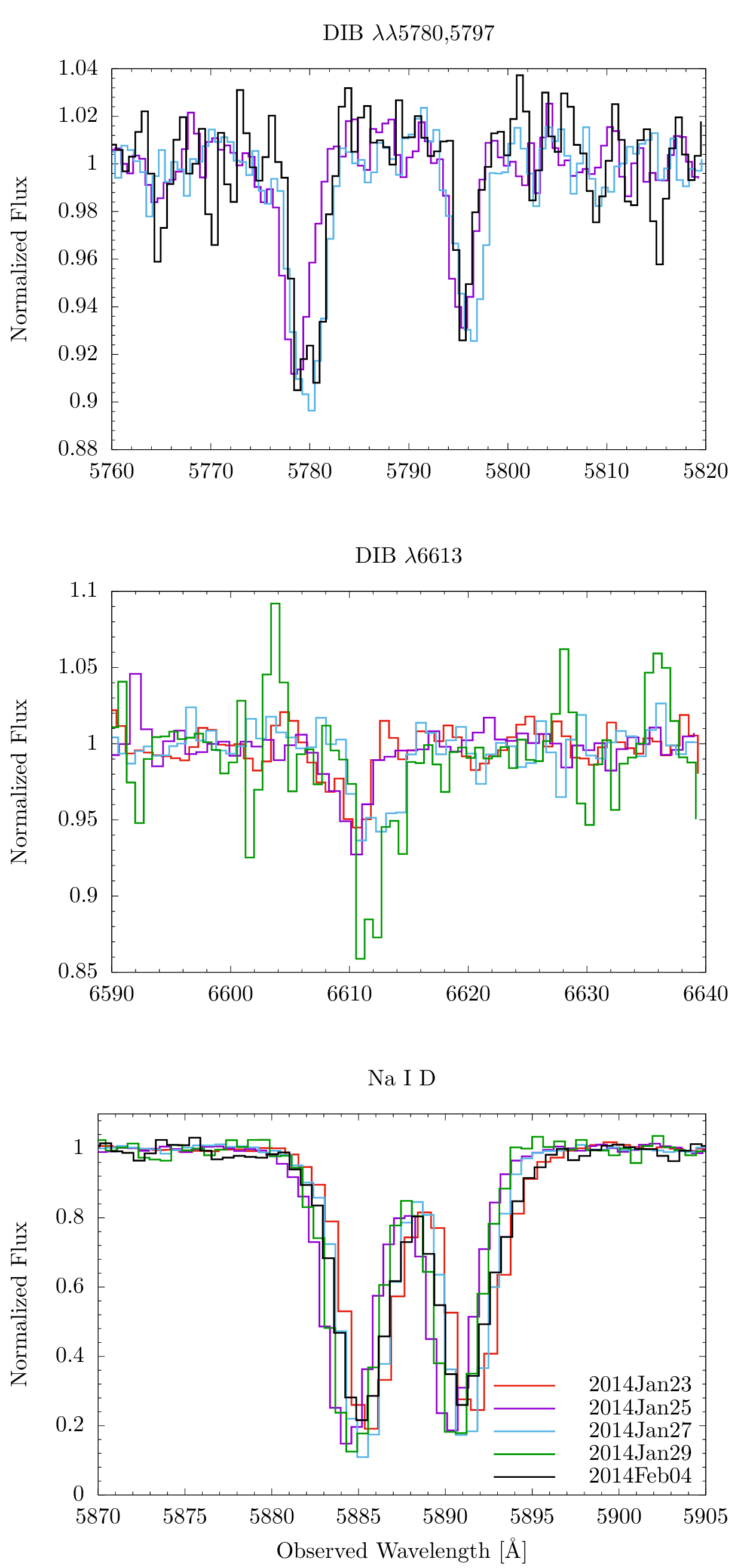}
\caption{The DIB $\lambda\lambda$5780,5797, DIB $\lambda$6613, and Na {\sc i} D absorption features detected in some of our intermediate-resolution spectra.}
\end{figure*}

\clearpage
\begin{figure*}
\centering{\includegraphics[width=\textwidth]{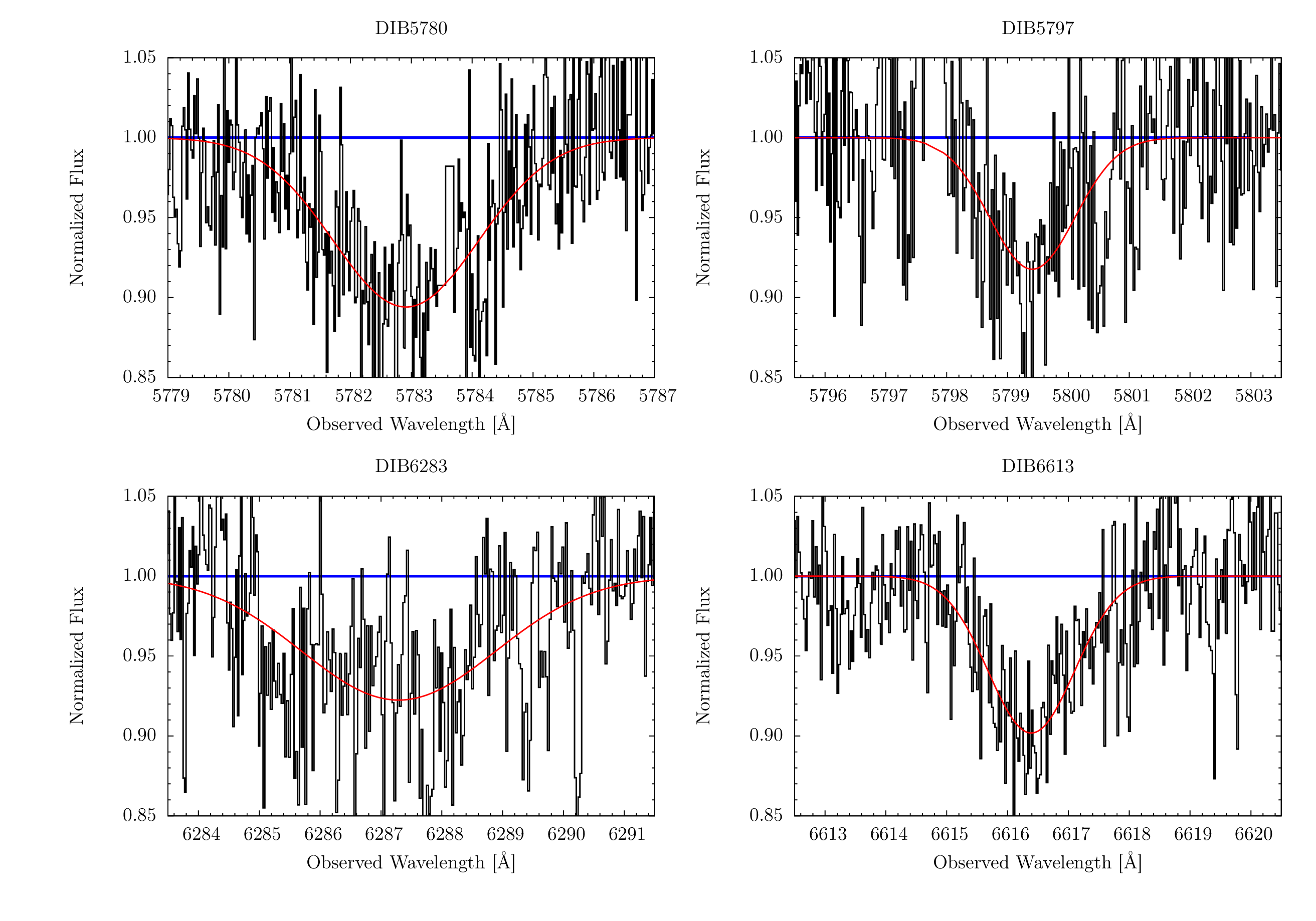}}
\caption{Normalized spectra of four detected DIB features. Black lines are our spectra, and red lines are Gaussian fits to the DIB features.}
\end{figure*}

\clearpage
\begin{figure*}
\includegraphics[width=\textwidth]{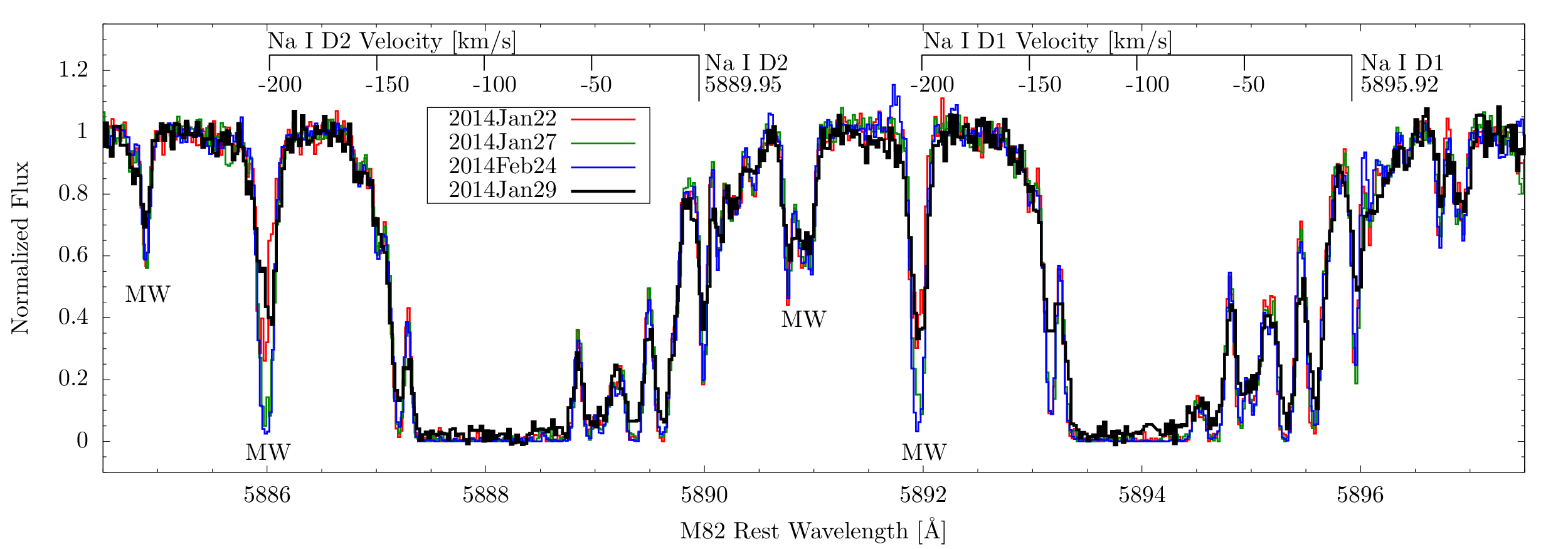}
\caption{Normalized Na {\sc i} D doublet of SN 2014J in our high-resolution spectrum taken on Jan. 29 2014 (in black). Spectra obtained by G15 on Jan. 22, Jan. 27, and Feb. 24 2014 are overplotted as red, green, and blue lines, respectively. Velocities in the rest frame of M82 are marked. Absorptions due to the Milky Way are also labeled.}
\end{figure*}

\clearpage
\begin{figure*}
\includegraphics[width=\textwidth]{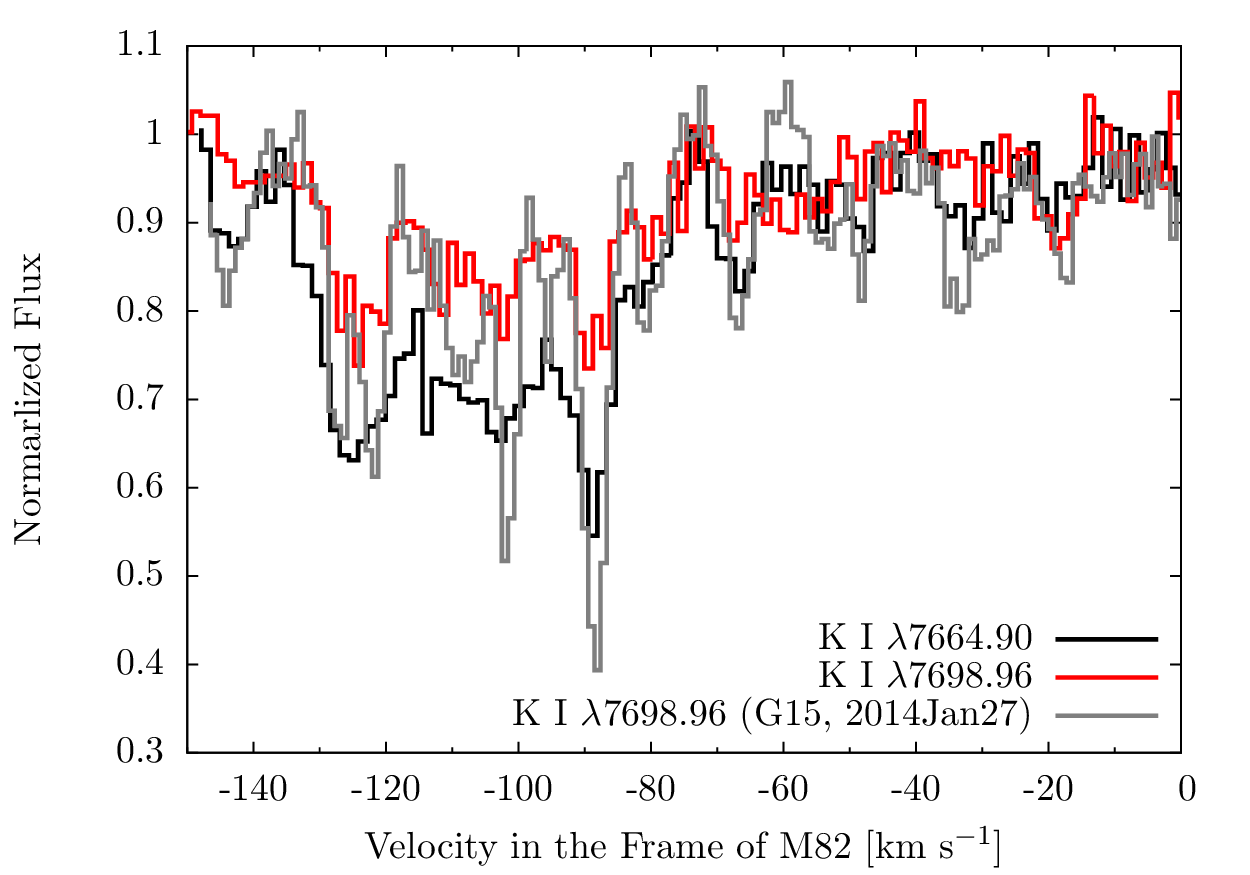}
\caption{Normalized K {\sc i} $\lambda$ 7665 (black) and K {\sc i} $\lambda$ 7699 (red) lines of SN 2014J in the high resolution spectrum taken with the 2.16-m telescope on Jan. 29 2014. The K {\sc i} 7699 from the spectrum taken by Graham et al. (2015, grey) is overplotted for comparison.}
\end{figure*}

\clearpage
\begin{figure*}
\includegraphics[width=\textwidth]{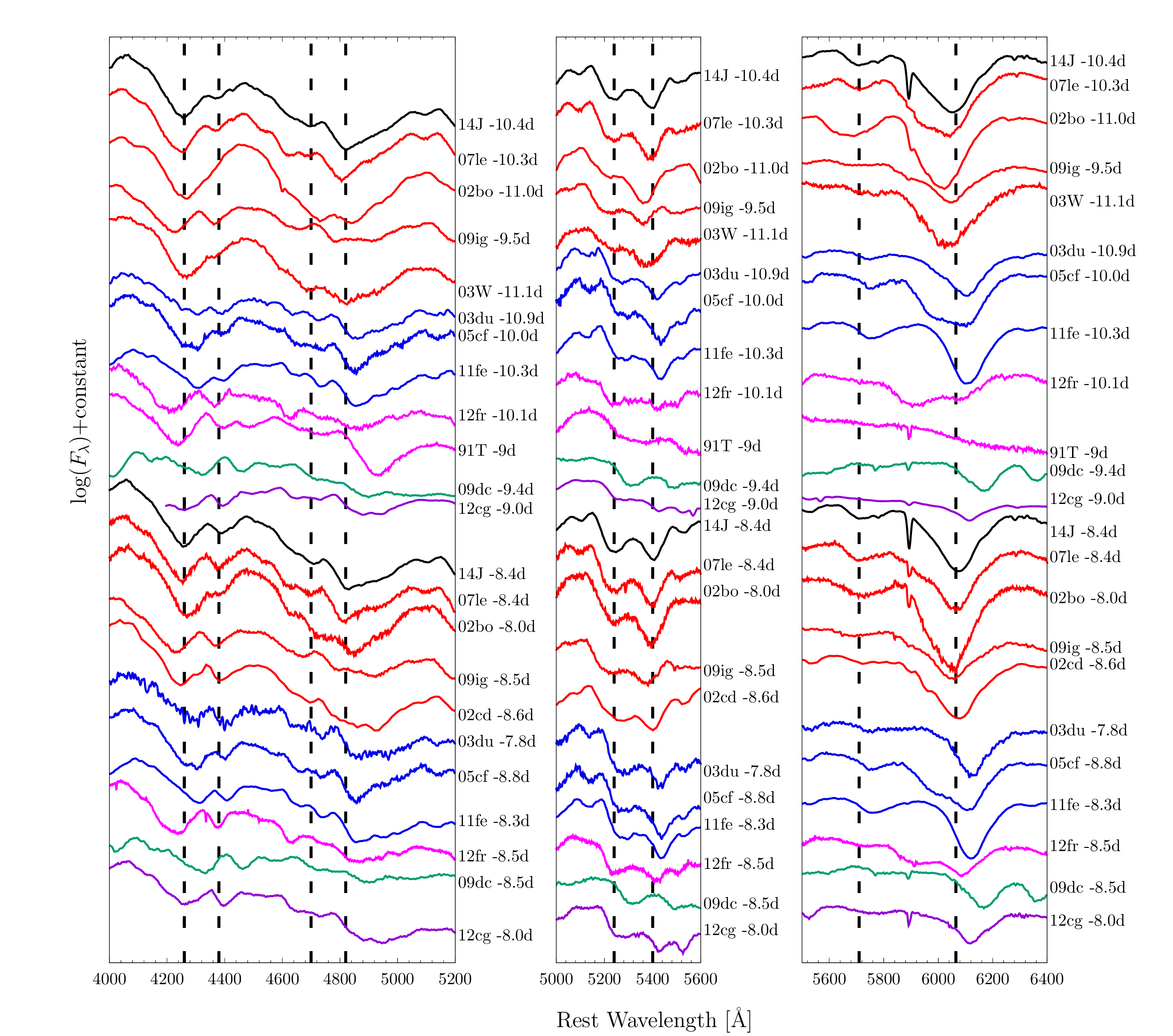}
\caption{Spectral evolution of SN 2014J and some comparison SNe in the wavelength ranges of 4000 - 5200 \AA, 5000 - 5600 \AA, and 5500 - 6400 \AA, at $\sim -$10 d. Vertical dash lines are over-plotted to mark the absorption troughs in SN 2014J. Red, blue, magenta spectra represent those of the HV, NV, 91T-like subtypes, respectively. Green and purple lines are for SNe 2009dc (Super-Chandrasekhar Candidate) and 2012cg (99aa-like). Spectra of SNe 1991T (Filippenko et al. 1992); 2002bo (Benetti et al. 2004); 2002cd (B12); 2003W (B12); 2003du (Stanishev et al. 2007); 2005cf (Wang et al. 2009b); 2007le (B12); 2009dc (Taubenberger et al. 2011); 2009ig (Marion et al. 2013);  2011fe (Pereira et al. 2013); 2012cg (Silverman et al. 2012b; Marion et al.2016); 2012fr (Childress et al. 2013) are used.}
\end{figure*}

\clearpage
\begin{figure*}
\includegraphics[width=\textwidth]{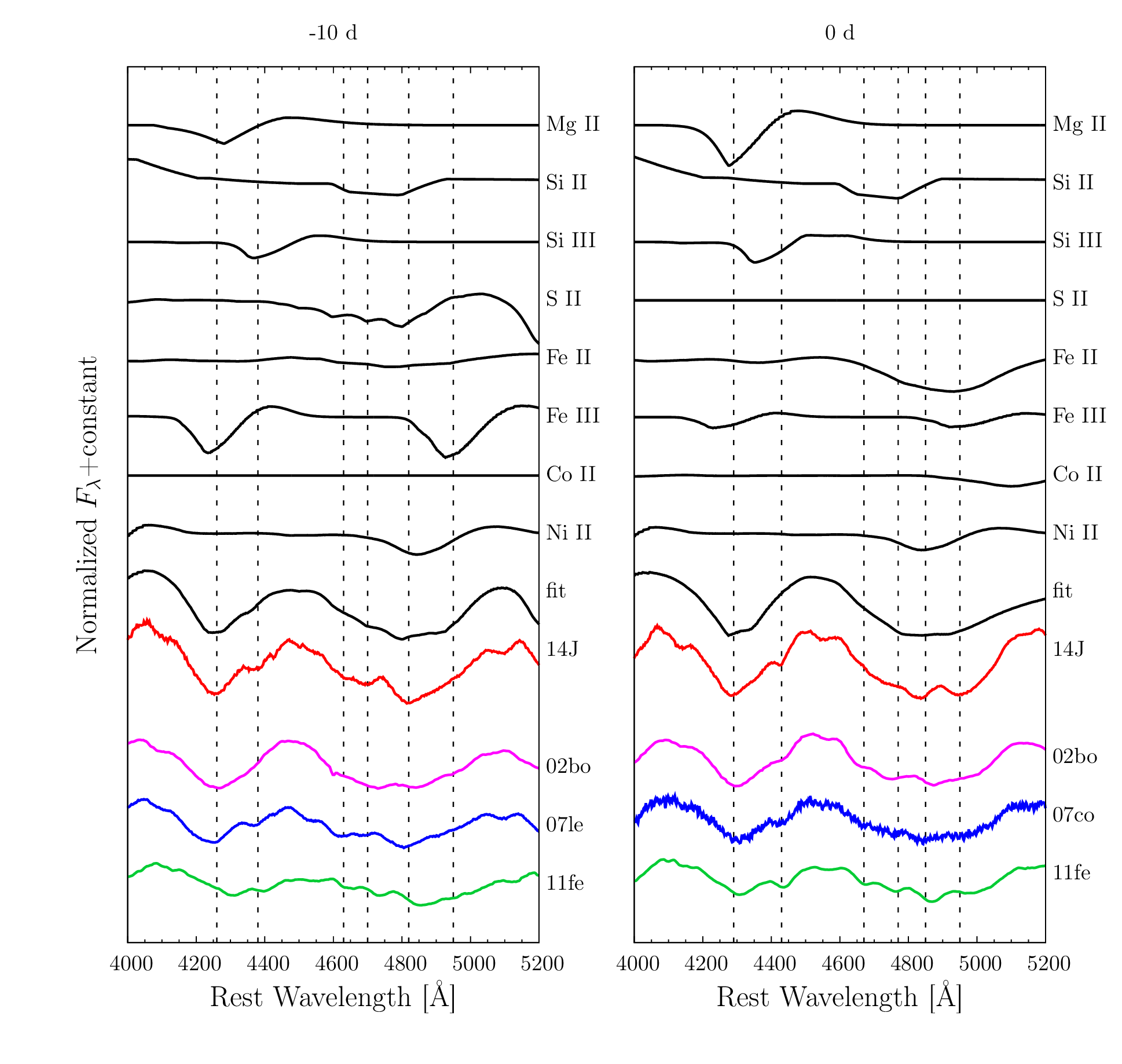}
\caption{\texttt{SYNAPPS} fit to the spectra at $-$10.4 d and  +0.6 d of SN 2014J from 4000 \AA\ to 5200 \AA. IMEs (Mg {\sc ii}, Si {\sc ii}, Si {\sc iii}, S {\sc ii}) and IGEs (Fe {\sc ii}, Fe {\sc iii}, Co {\sc ii}, Ni {\sc ii}) are used. Spectra of SNe 2002bo, 2007co, 2011fe are also plotted for comparison, and the absorption features are marked by the dashed lines.}
\end{figure*}

\clearpage
\begin{figure*}
\includegraphics[width=\textwidth]{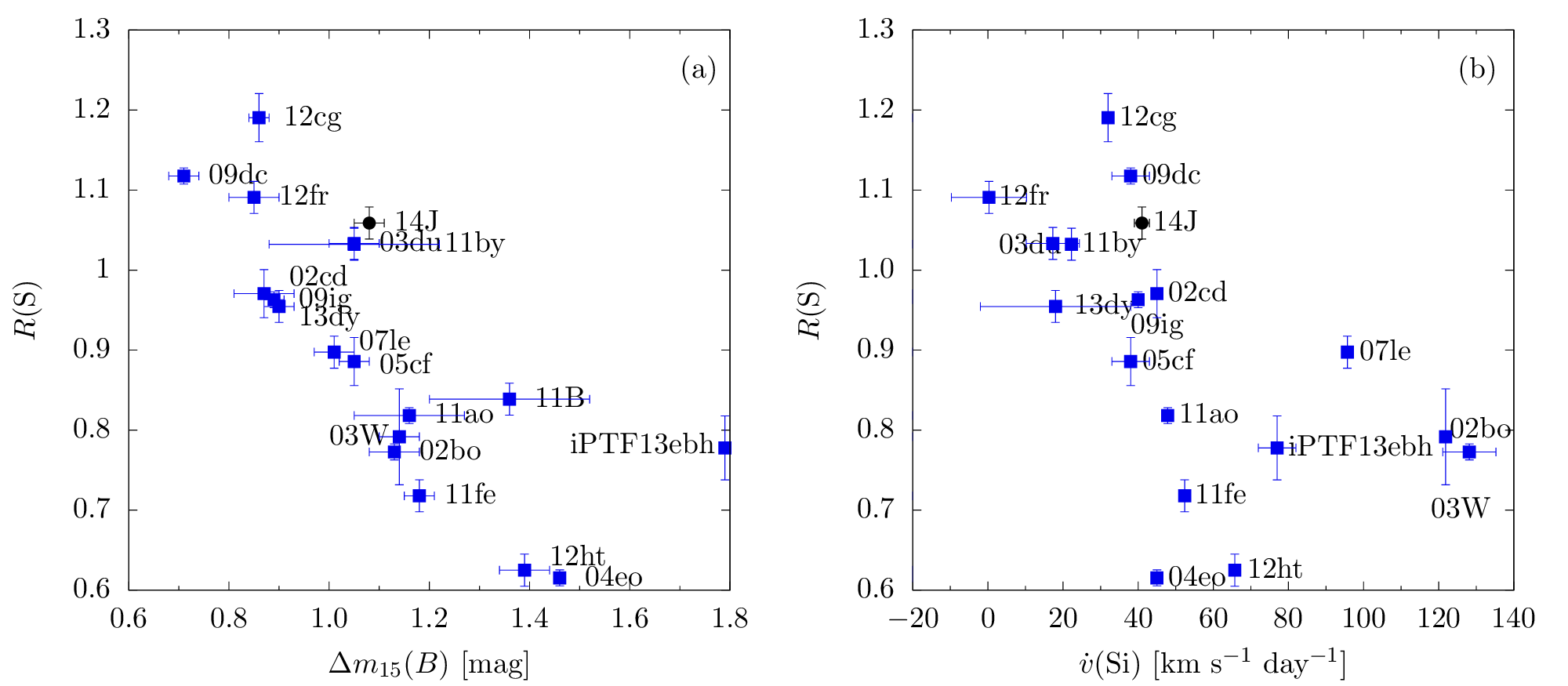}
\caption{(a) The ratio of S~{\sc ii} $\lambda$5468 and S~{\sc ii} $\lambda$5640 absorptions, $R$(S) versus $\Delta m_{15}(B)$ for some well-observed SNe Ia. (b) The correlation of $R$(S) versus the velocity gradient $\dot{v}$(Si) for the same sample.}
\end{figure*}
\clearpage
\begin{table}
	\centering
	\caption{Journal of Spectroscopic Observations of SN 2014J}
	\begin{tabular}{cccccc}
		\hline
		UT Date & JD$^a$ & Phase$^b$ & Exp.(s) & Telescope + Intrument & Range \\
		\hline
		2014-01-22 & 6679.3 & $-$10.4 & 900 & YNAO 2.4~m+YFOSC (G3) & 3400-9000 \\
		2014-01-22 & 6679.3 & $-$10.4 & 1800 & YNAO 2.4~m+YFOSC (E13) & 5800-8500 \\
		2014-01-22 & 6679.9 & $-$9.8 & 240 & 1.22~m+B\&C & 3300-7900 \\
		2014-01-23 & 6680.4 & $-$9.4 & 2400 & YNAO 2.4~m+YFOSC (G10+E13) & 3900-9000 \\
		2014-01-24 & 6681.3 & $-$8.4 & 900 & YNAO 2.4~m+YFOSC (G3) & 3500-9100 \\
		2014-01-25 & 6682.4 & $-$7.3 & 1500 & YNAO 2.4~m+YFOSC (G10+E13) & 3860-9200 \\
		2014-01-25 & 6682.8 & $-$6.9 & 600 & 1.82~m+AFOSC & 3550-10100 \\
		2014-01-25 & 6683.0 & $-$6.7 & 240 & TNG+LRS & 3600-10200 \\
		2014-01-26 & 6683.4 & $-$6.3 & 600 & YNAO 2.4~m+YFOSC (G3) & 3500-9100 \\
		2014-01-27 & 6684.2 & $-$5.5 & 360 & BAO 2.16 m+BFOSC & 3800-8800 \\
		2014-01-27 & 6684.4 & $-$5.4 & 1500 & YNAO 2.4~m+YFOSC (G10+E9) & 4000-8900 \\
		2014-01-28 & 6685.4 & $-$4.4 & 600 & YNAO 2.4~m+YFOSC (G3) & 3500-9100 \\
		2014-01-29 & 6686.2 & $-$3.5 & 300 & 1.22~m+B\&C & 3300-7900 \\
		2014-01-29 & 6686.4 & $-$3.4 & 1800 & YNAO 2.4~m+YFOSC (G10+E9) & 4000-9200 \\
		2014-01-30 & 6687.4 & $-$2.4 & 600 & YNAO 2.4~m+YFOSC (G3) & 3500-9100 \\
		2014-01-31 & 6688.4 & $-$1.3 & 600 & YNAO 2.4~m+YFOSC (G3) & 3500-9100 \\
		2014-02-02 & 6690.3 & +0.6 & 600 & YNAO 2.4~m+YFOSC (G3) & 3500-9100 \\
		2014-02-04 & 6692.4 & +2.7 & 2400 & YNAO 2.4~m+YFOSC (G10+E13) & 4000-9400 \\
		2014-02-05 & 6693.4 & +3.6 & 600 & YNAO 2.4~m+YFOSC (G3) & 3500-9100 \\
		2014-02-07 & 6695.0 & +5.3 & 1800 & 1.22~m+B\&C & 3200-7900 \\
		2014-02-09 & 6698.0 & +8.2 & 960 & 1.22~m+B\&C & 3300-7900 \\
		2014-02-10 & 6698.4 & +8.7 & 300 & YNAO 2.4~m+YFOSC (G3) & 3500-9100 \\
		2014-02-12 & 6700.8 & +11.1 & 1680 & 1.22~m+B\&C & 3400-7900 \\
		2014-02-14 & 6702.8 & +13.1 & 840 & 1.22~m+B\&C & 3400-7900 \\
		2014-02-18 & 6706.1 & +16.4 & 1200 & 1.22~m+B\&C & 3300-7900 \\
		2014-02-19 & 6707.3 & +17.5 & 480 & YNAO 2.4~m+YFOSC (G3) & 3500-9100 \\
		2014-02-24 & 6712.2 & +22.5 & 300 & YNAO 2.4~m+YFOSC (G3) & 3500-9100 \\
		2014-02-27 & 6715.0 & +25.3 & 120 & TNG+LRS & 3500-10300 \\
		2014-03-02 & 6718.3 & +28.5 & 480 & YNAO 2.4~m+YFOSC (G3) & 3500-9100 \\
		2014-03-06 & 6722.8 & +33.1 & 4500 & 1.22~m+B\&C & 3300-7900 \\
		2014-03-07 & 6724.1 & +34.4 & 3600 & BAO 2.16 m+OMR & 3900-9200 \\
		2014-03-11 & 6727.0 & +37.3 & 600 & 1.82~m+AFOSC & 3900-10100 \\
		2014-03-12 & 6728.8 & +39.1 & 3600 & 1.22~m+B\&C & 3300-7900 \\
		2014-03-14 & 6730.2 & +40.5 & 600 & YNAO 2.4~m+YFOSC (G3) & 3500-9100 \\
		2014-03-20 & 6736.9 & +47.1 & 3600 & 1.22~m+B\&C & 3300-7900 \\
		2014-03-21 & 6738.1 & +48.4 & 600 & BAO 2.16 m+BFOSC & 4200-8700 \\
		2014-04-01 & 6748.0 & +58.3 & 600 & 1.82~m+AFOSC & 3900-9300 \\
		2014-04-07 & 6754.9 & +65.1 & 1200 & 1.82~m+AFOSC & 3900-9300 \\
		2014-04-16 & 6763.9 & +74.1 & 4800 & 1.22~m+B\&C & 3300-7900 \\
		2014-04-18 & 6765.2 & +75.5 & 1200 & YNAO 2.4~m+YFOSC (G3) & 3900-9100 \\
		2014-04-24 & 6771.5 & +81.8 & 3600 & BAO 2.16 m+BFOSC & 3800-8700 \\
		2014-04-25 & 6772.5 & +82.8 & 1500 & YNAO 2.4~m+YFOSC (G3) & 3900-9100 \\
		2014-05-03 & 6780.9 & +91.1 & 5400 & 1.22~m+B\&C & 3300-7900 \\
		2014-05-08 & 6785.8 & +96.1 & 1200 & 1.22~m+B\&C & 3800-7900 \\
		2014-05-19 & 6796.6 & +106.8 & 1200 & BAO 2.16 m+BFOSC & 3800-8700 \\
		2014-05-22 & 6799.0 & +109.3 & 1200 & 1.22~m+B\&C & 4000-7900 \\
		2014-05-22 & 6799.5 & +109.8 & 1200 & BAO 2.16 m+BFOSC & 3800-8700 \\
		2014-06-10 & 6818.9 & +129.2 & 5400 & 1.82~m+AFOSC & 3400-9300 \\
		2014-11-25 & 6986.0 & +296.3 & 1800 & 1.82~m+AFOSC & 4000-8100 \\
		2014-11-25 & 6986.4 & +296.7 & 2274 & YNAO 2.4~m+YFOSC (G3) & 4000-9100 \\
		2014-12-20 & 7011.1 & +321.4 & 2400 & 1.82~m+AFOSC & 4200-8100 \\
		2015-03-10 & 7091.9 & +402.2 & 5400 & 1.82~m+AFOSC & 4000-8000 \\
		2015-03-12 & 7093.9 & +404.2 & 5400 & 1.82~m+AFOSC & 4000-8000 \\
		2015-04-05 & 7117.9 & +428.2 & 1800 & GTC+OSIRIS+grism R300B & 4000-10000 \\
		2015-05-20 & 7162.9 & +473.2 & 3000 & GTC+OSIRIS+grism R300B & 4000-10000 \\
		\hline
		\multicolumn{3}{l}{$^a$ The Julian Date is subtracted by 245,0000.5}\\
		\multicolumn{3}{l}{$^b$ Relative to the epoch of $B$-band maximum (MJD 56689.74 $\pm$ 0.13).}\\
	\end{tabular}
\end{table}		

\clearpage
\begin{table}
	\centering
	\caption{Parameters of SN 2014J and Some Comparison SNe Ia}
	\begin{tabular}{ccccccc}
		\hline
		SN & $\Delta m_{15}(B)_{true}$ [mag] & Subtype$^a$ & $v$ [km s$^{-1}$] & $\dot{v}$ [km s$^{-1}$ day$^{-1}$] & $B_{\rm max}-V_{\rm max}^b$ [mag] & Reference\\
		\hline
		2002bo & 1.17$\pm$0.05 & BL, HV & 13121 & 128 & 0.46$\pm$0.14 & c,d \\
		2003du & 1.02$\pm$0.05 & CN, NV & 10522 & 17 & $-$0.08$\pm$0.03 & c,e \\
		2005cf & 1.07$\pm $0.03 & CN, NV & 10340 & 62 & 0.07$\pm$0.03 & c,f \\
		2007co & 1.16$\pm $0.03 & BL, HV & 12024 & 99 & $-$0.08$\pm$0.04 & c \\
		2007le & 1.06$\pm $0.03 & BL, HV & 12780 & 93 & 0.37$\pm$0.04 & c,g,h \\
		2009ig & 0.91$\pm$0.07 & CN, HV & 13400 & 40 & -0.04$\pm$0.17 & i \\
		2011fe & 1.18$\pm$0.03 & CN, NV & 10400 & 52.4 & $-$0.03$\pm$0.04 & j \\
		2014J & 1.08$\pm$0.03 & BL, HV & 12200 & 41 & 1.28$\pm$0.03 & k,l \\
		\hline
		\multicolumn{3}{l}{$^a$ In Branch's and Wang's classification schemes respectively. References are in the text.}\\
		\multicolumn{3}{l}{$^b$ The colors have been corrected for the Galactic reddening.}\\
		\multicolumn{3}{l}{$^c$ Blondin et al. (2012)}\\
		\multicolumn{3}{l}{$^d$ Benetti et al. (2004)}\\
		\multicolumn{3}{l}{$^e$ Stanishev et al. (2007)}\\
		\multicolumn{3}{l}{$^f$ Wang et al. (2009b)}\\
		\multicolumn{3}{l}{$^g$ Silverman et al. (2012a)}\\
		\multicolumn{3}{l}{$^h$ Wang et al. (2009a)}\\
		\multicolumn{3}{l}{$^i$ Marion et al. (2013)}\\
		\multicolumn{3}{l}{$^j$ Zhang et al. (2016)}\\
		\multicolumn{3}{l}{$^k$ Srivastav et al. (2016)}\\
		\multicolumn{3}{l}{$^l$ This paper}\\
	\end{tabular}
\end{table}

\begin{table}
	\centering
	\caption{Equivalent Width ($EW$) of Na {\sc i} D1 and D2 Absorptions}
	\begin{tabular}{ccc}
		\hline
		Phase & $EW$(D1) [\AA] & $EW$(D2) [\AA] \\
		\hline
		$-$9.4 & 2.89 & 2.73 \\
		$-$7.3 & 3.00 & 2.69 \\
		$-$5.4 & 3.03 & 2.64 \\
		$-$3.4 & 3.02 & 2.68 \\
		$+$2.7 & 2.91 & 2.81 \\
		\hline
	\end{tabular}
\end{table}

\begin{table}
	\centering
	\caption{Gaussian fit to the parameters of DIBs}
	\begin{tabular}{cccccc}
		\hline
		Source & FWHM (\AA) & $EW$ (\AA) & A$_{v}$$^a$ & A$_{v}$$^b$ & A$_{v}$$^c$ \\
		\hline
		DIB 5780 & 2.78 $\pm$ 0.33 & 0.32 $\pm$ 0.05 & 1.7 $\pm$ 0.9 & 1.9 $\pm$ 0.3 & 1.8 \\
		DIB 5797 & 1.63 $\pm$ 0.45 & 0.14 $\pm$ 0.05 & \nd & 2.4 $\pm$ 0.9 & 2.1 \\
		DIB 6283 & 3.87 $\pm$ 0.74 & 0.32 $\pm$ 0.08 & \nd & 0.7 $\pm$ 0.2 & 0.6\\
		DIB 6613 & 1.69 $\pm$ 0.24 & 0.18 $\pm$ 0.03 & \nd & 2.6 $\pm$ 0.4 & 2.4 \\
		\hline
		\multicolumn{3}{l}{$^a$ Using the relationship in Phillips et al. (2013).}\\
		\multicolumn{3}{l}{$^b$ Using the relationship in Friedman et al. (2011).}\\		
		\multicolumn{3}{l}{$^c$ Using the relationship in Welty et al. (2014).}\\		
	\end{tabular}
\end{table}

\bsp	
\label{lastpage}
\end{document}